\documentclass[aps,prb,groupedaddress]{revtex4}
\usepackage[dvips]{graphics}
\usepackage{vmargin}
\setpapersize{USletter}
\setmarginsrb{1in}{1in}{1in}{1in}{0in}{0in}{0in}{0in}

\begin{document}
\title{Propagation of a quantum particle through two dimensional percolating systems}
\author{Md Fhokrul Islam}
\author{Hisao Nakanishi}
\affiliation{Department of Physics, Purdue University, West Lafayette, IN 47907}
\date{\today}

\begin{abstract}
We study quantum percolation which is described by a tight-binding Hamiltonian
containing only off-diagonal hopping terms that are generally in quenched binary
disorder (zero or one). In such a system, transmission of a quantum particle is
determined by the disorder and interference effects, leading to interesting sharp
features in conductance as the energy, disorder, and boundary conditions are
varied. To aid understanding of this phenomenon, we develop a visualization method
whereby the progression of a wave packet entering the cluster through a lead on one
side and exiting from another lead on the other side can be tracked dynamically.
Using this method, we investigate the evolution of the wave packet through
transmission and reflection under clusters of various disorder, sizes, boundary
conditions, and energies. Our results indicate the existence of two different kinds
of localized regimes, namely exponential and power law localization, depending on
the amount of disorder. Our study further suggests that there may be a
delocalized state in the 2D quantum percolation system at very low disorder.
These results are based on a finite size scaling analysis of the systems of size
up to $70 \times 70$ (containing 4900 sites) on the square lattice.
\end{abstract}

\maketitle

\section{Introduction}
The transmittance of a classical particle through a disordered system depends solely
on the availability of a spanning path through the system. This criterion, however,
must be modified generally for a quantum particle since interference and tunneling
effects must be included for a quantum system.  In particular, the presence of a
spanning path does not by itself ensure the transmittance of a quantum particle
because of the interference effect. Even for a completely ordered system, a quantum
particle may exhibit zero transmission depending on the details such as the energy
of the particle \cite{cuansing:04}. Our quantum percolation system includes
the interference effect fully but does not include any of the tunneling effect;
thus we expect a higher connectivity in underlying geometry to be required for
non-zero transmission compared to its classical counterpart.

A major motivation for studying such a system is the question of whether a localized-
to-delocalized (or perhaps, metal-to-insulator) transition exists in a two-dimensional
(2D) system. The Anderson model and the quantum percolation model are
two of the more common theoretical models that are used to study the transport
properties of disordered systems. While the literature on both models agree on
the existence of such a transition in three dimensions,\cite{souk:87,kosl:91,berk:96},
the same question for quantum percolation in two dimensions appears to have remained
a subject of controversy for over two decades. Based on the scaling theory of Abrahams et al
\cite{abrahams:79}, it was widely believed that there can be no metal-to-insulator
transition in 2D universally in the absence of a magnetic field or interactions for
any amount of disorder.\cite{goldenfeld} However, whether this theory also applies to
quantum percolation has been debated in recent years.

In the mean time, experiments performed in early 1980's on different 2D systems
\cite{exp:1,exp:2,exp:3} confirmed the scaling theory predictions. However, a
number of experiments that appeared more recently seem to suggest that a metallic
state may be possible in two dimensions. For reviews of these experiments, see
Abrahams {\it et al.} \cite{abrahams:01} and references therein. In this work, we
do not address the issues of these experiments, but rather concentrate on the
formally much simpler quantum percolation model which has neither magnetic field
nor interactions but contains binary disorder with infinite barriers at randomly
diluted sites.

We study the behavior of 2D quantum percolation systems by tracking how a wave
packet, representing a quantum particle, evolves with time as it enters a diluted
2D system through a 1D lead. Even restricting attention to quantum percolation
which lacks many effects that are expected to play important roles in metal-insulator
transitions, there is a long-standing controversy as to the presence or absence of
an extended state and of a phase transition between the prevalent localized state
and a more elusive extended state in two-dimensions. On one hand, some studies such
as those made using the dlog Pad\'{e} approximation method \cite{daboul:00}, real
space renormalization method \cite{odagaki84}, and the inverse participation ratio
\cite{srivastava84} found a transition from exponentially localized states to
non-exponentially localized states for a range of site concentrations between
$0.73 \leq p_q \leq 0.87$ on the square lattice. So did a study of energy level
statistics \cite{letz:99}, one of the spread of a wave packet initially localized
at a site \cite{nazareno:02}, and one of a transfer matrix \cite{eilmes:01}, where
the nature of the delocalized state remained not fully understood. On the other
hand, studies such as the scaling work based on numerical calculation of the
conductance \cite{haldas02}, the investigation of vibration-diffusion analogy
\cite{bunde98}, finite-size scaling analysis and transfer matrix methods
\cite{soukoulis91}, and vector recursion technique \cite{mookerjee95} found no
evidence of a transition.  A study by Inui {\it et al.} \cite{inui94} found all
states to be localized except for those with particle energies at the middle of
the band and when the underlying lattice is bipartite, such as a square lattice.
More recently, Cuansing and Nakanishi (private communication) used an approach first
suggested by Daboul {\it at al.} \cite{daboul:00} to calculate conductance directly
for clusters of several hundred sites and, extrapolating those results by finite-
size scaling, suggested that delocalized states exist and thus a transition would
have to exist as well.

To aid understanding of transport of a quantum particle in 2D systems, we have
developed a new dynamical visualization method. It is based on usual scattering
process where we let a particle, described by a Gaussian wave packet, enter into
the system (sometimes called a {\em cluster}) through an input lead and track
how it propagates through the system and exits from the other side of the cluster
through the output lead. Thus by determining the transmission probability we can
acquire valuable information about the cluster. The method is very useful to study
the characteristics of ordered as well as disordered clusters. It clearly demonstrates
the effect of interference on propagation which is a unique feature of a quantum
system. A great utility of this method is that we can visualize the progression of
the wave packet in real time and investigate the nature of localization of a particle
in 2D quantum percolation system.

We have used the tool we have developed to study behavior of a particle both in
square and triangular lattices. However, most of the work presented in this paper
is for a 2D system realized on the square lattice. In Section II, we discuss the
method we have developed to study 2D systems. In Sections III and IV we present the
application of this method to ordered and disordered quantum percolation systems
respectively. And finally we present our conclusion in Section V.

\section{The Model and Numerical approach}

We study quantum percolation that is described by

\begin{equation}
H = \sum_{<ij>} V_{ij}|i\rangle {\langle}j| + h.c \label{eq1}
\end{equation}

where $|i>$ and $|j>$ are tight binding basis functions at sites $i$ and $j$,
respectively and $V_{ij}$ is the hopping matrix element which is 1 if $i$ and $j$
are nearest neighbors, otherwise zero. We realized this model on both square and
triangular lattices that can have at most 4 and 6 nearest neighbors respectively.
In ordered cases, all lattice sites are available and included in the sum, while
in disordered cases, we randomly dilute (or remove) a certain fraction of the
lattice sites and consider them not available (and excluded from the sum).

To study the propagation of a quantum particle we connect two 1D leads, one as
input and the other one as output lead, to the 2D cluster, and construct a Gaussian
wave packet in the input chain that represents the incident particle. The usual
time dependent Schr\"{o}dinger equation is then solved for this system.

\begin{equation}
i\frac{\partial\psi (r,t)}{\partial t} = H\psi \label{eq2}
\end{equation}
Here,
\begin{equation}
\psi(r,t=0) = (const.)\cdot\mbox{e}^{(r-r_{0})^{2}/(4s^{2})}\cdot\mbox{e}^{-ik_{0}r}
\label{eq3}
\end{equation}

(on input chain only) and $s$ is the width of the wave packet, $k_{0}$ is the central
wave vector of the wave packet and $r_{0}$ is the location of the center of the wave
packet in the input chain. For all simulations in this work we construct the wave
packets with $s$ which corresponds to 4 lattice spacings and $r_{0}$ at the middle
of the input chain. These choices assure that $k_0$ is a well-defined mean wave
vector of the incoming wave packet.

Allowing this wave packet to evolve with time under this Hamiltonian we can study
its transmission characteristics. For simulation purposes, Eq.~\ref{eq2} is converted
to a discrete form as follows \cite{nakanishi:05}. We write the wave function in
terms of its real and imaginary parts and upon substitution back into Eq.~\ref{eq2}
this gives two coupled equations:

\begin{eqnarray}
\psi(r,t) = R(r,t) + iI(r,t) \\
\frac{\partial R(r,t)}{\partial t} = H\cdot I(r,t) \\
\frac{\partial I(r,t)}{\partial t} = -H\cdot R(r,t)
\label{eq4}
\end{eqnarray}

We then apply a leap-frog method to arrive at the discrete forms of
these two equations,

\begin{eqnarray}
I(r,t+\frac{1}{2})\approx I(r,t-\frac{1}{2})- \triangle t\cdot H \cdot R(r,t)\\
R(r,t+1) \approx R(r,t)+ \triangle t \cdot H \cdot I(r,t+\frac{1}{2})
\label{eq7}
\end{eqnarray}

We have used MATLAB graphics tools to visualize this dynamical process.

\section{Propagation through a completely ordered system}
\subsection{Square Lattice}

\begin{figure}[h]
{\resizebox{3.3in}{2.9in}{\includegraphics{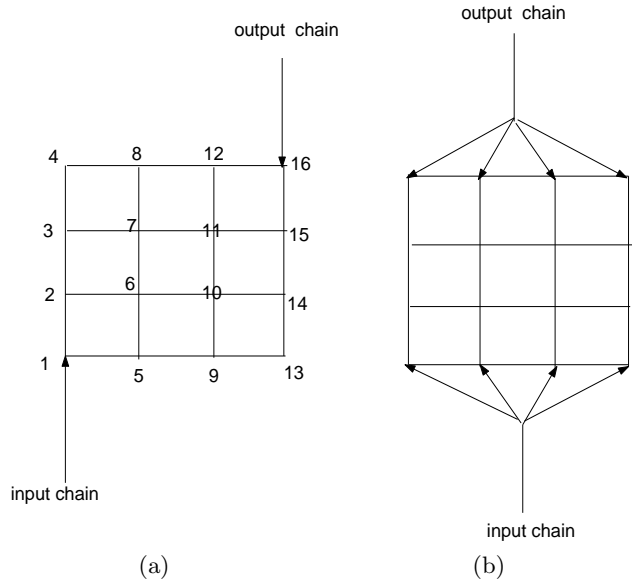}}}\\
(a) \hspace{1.5in} (b)
\caption{$4 \times 4$ Square Lattice: (a) point to point connection and (b) busbar
        type connection. The numbers are the labels of the lattice points of the
        cluster used to construct the Hamiltonian (Eq.~\ref{eq1}). The same sequence
        of labeling is used for all sizes of the clusters used in this work.}
\label{sq_lat}
\end{figure}

We have applied the method described in the previous section to investigate first
the behavior of a quantum particle in an ordered 2D system. In this context, {\em
ordered} means that all lattice sites of the system are available to the particle.
The cluster has been studied using two different types of the connection of the
leads. In the point-to-point contact, the input lead is connected to only one site
on the input side of the cluster and the output lead is connected to only one
lattice site on the opposite side of the cluster. In the busbar type contact, all
the lattice points on the input side of the cluster are connected to the input lead,
while all the lattice points on the output side of the cluster are connected to the
output lead (Fig.~\ref{sq_lat}).

\begin{figure}[h]
$\begin{array}{cc}
{\resizebox{3.2in}{2in}{\includegraphics{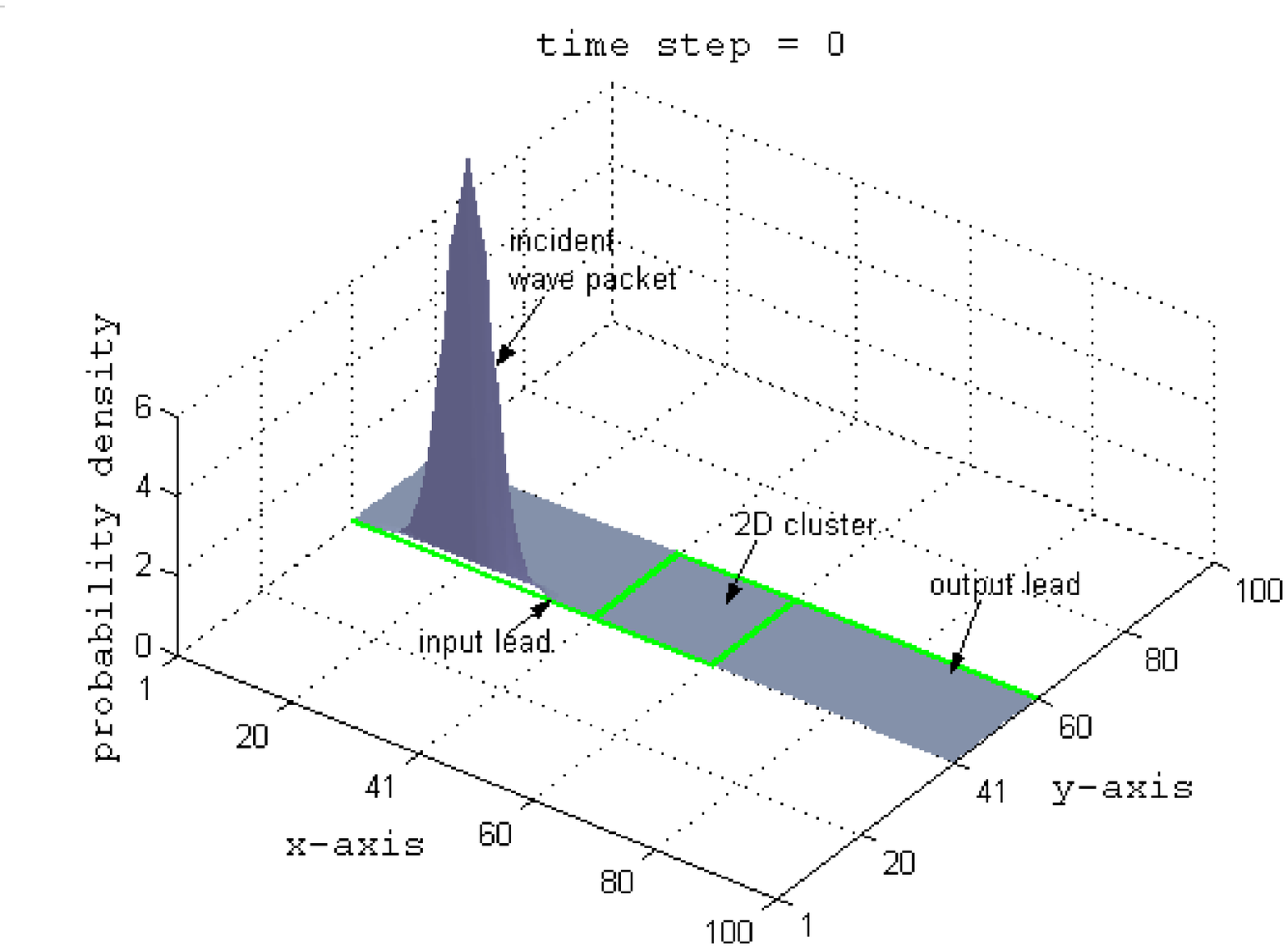}}} &
{\resizebox{3.2in}{2in}{\includegraphics{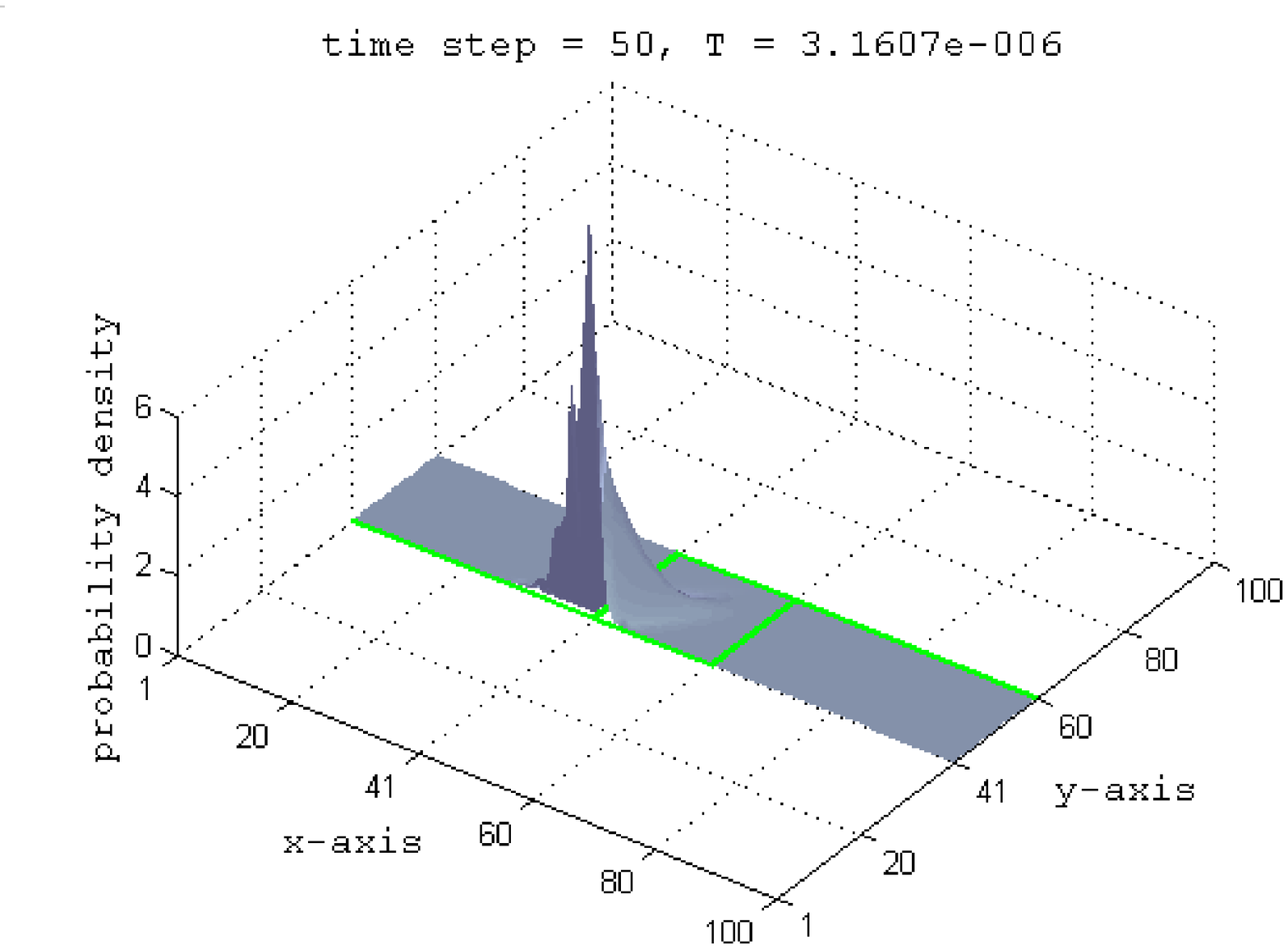}}} \\
(a) & (b)\\
{\resizebox{3.2in}{2in}{\includegraphics{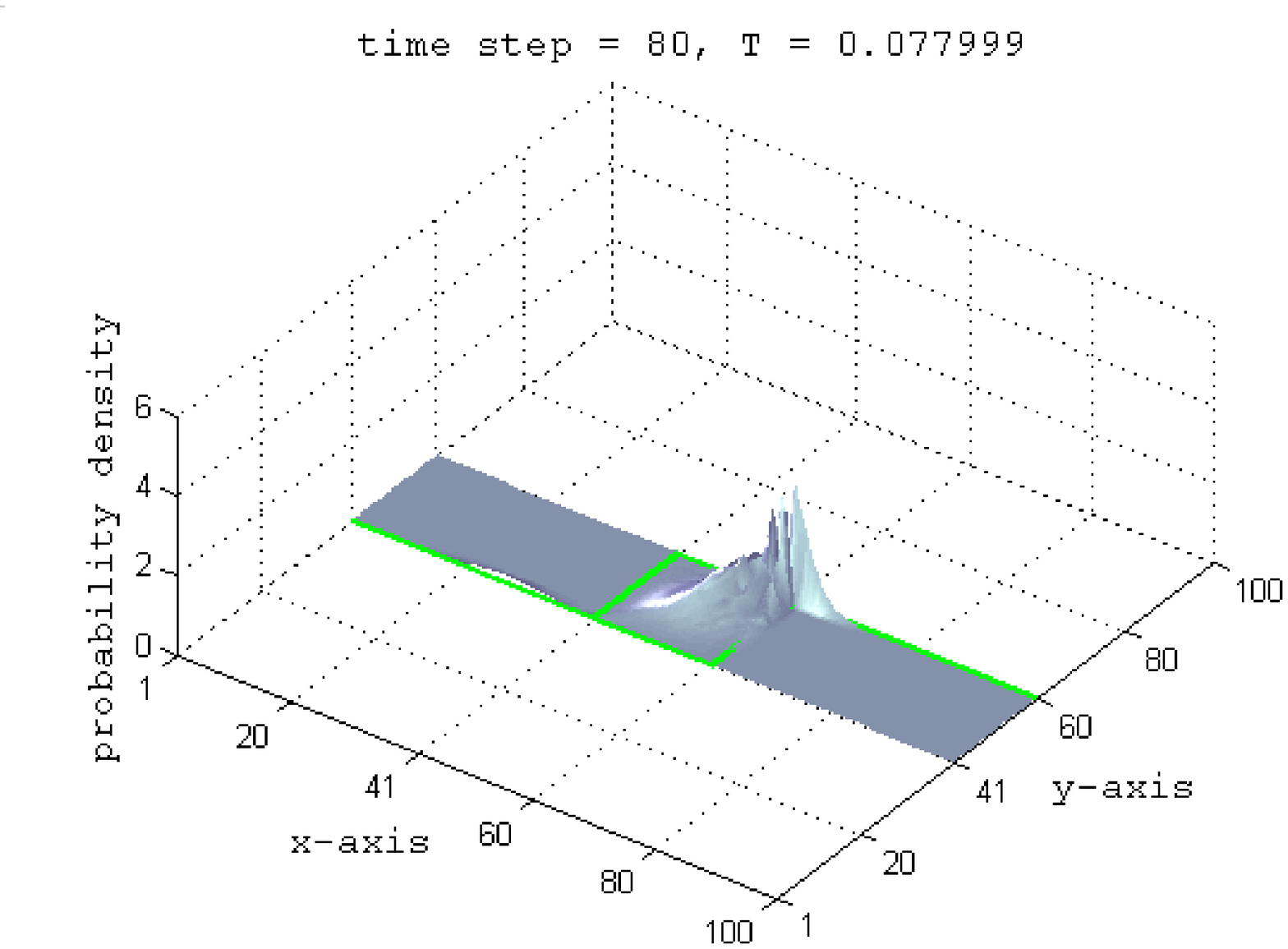}}} &
{\resizebox{3.2in}{2in}{\includegraphics{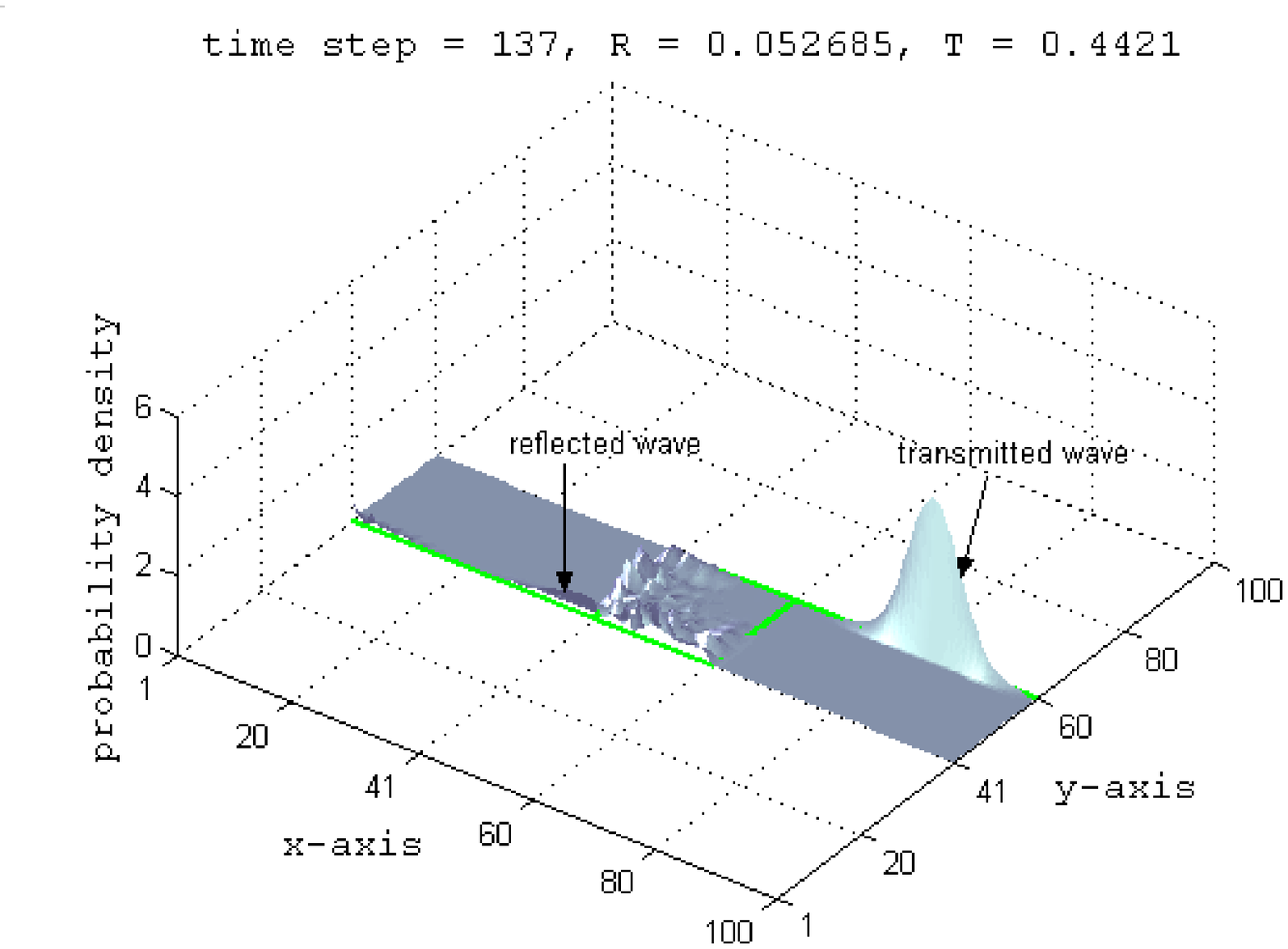}}} \\
(c) & (d)
\end{array}$

\caption{A typical time evolution of a wave packet through a cluster of size
        $20 \times 20$ with leads connected diagonally at the lower left and
        upper right corners of the cluster. The time step for each snapshot and
        corresponding reflection and transmission coefficients $R$ and $T$ are
        shown at the top of each snapshot.}
\label{p2p_diag}
\end{figure}

There are many possible ways one can arrange a point-to-point connection and
transmission is affected by the details of the leads' connection to the cluster.
One such arrangement is shown in Fig.~\ref{p2p_diag}, where a packet propagates
through diagonally connected leads. We generate a wave packet centered in the
middle of the input lead with a central wave vector $k_{0}$ = 4.3 (in units of
inverse lattice constant). As the wave packet spreads and propagates through the
cluster, it interferes with itself in many ways. This interference effect is
particularly large when the wave packet reaches the contact point with the output
lead (Fig.~\ref{p2p_diag}c). A part of the wave packet is reflected back to
the cluster from the contact and surrounding edges of the cluster, thereby reducing
transmission. The calculation of the transmission coefficient $T$ requires some
approximation in this approach. In most of what follows, we estimate $T$ by summing
the portion of the normalized wave packet that is on the output lead when the
leading edge of the wave packet has first reached the end of the output lead. Often
we can visually confirm the validity of this method by looking at the evolution of
the wave packet shape explicitly, but in addition, we have also first employed
increasing lead lengths to locate an optimal length where this approximation
yields stable values of $T$ and then used this optimal length in our later work.

Another important feature that is evident from Fig.~\ref{p2p_diag} is that,
unlike a classical particle, a quantum particle has a non-zero probability to be
observed at any part of the system at any instant of time. However, since we are
most interested in the behavior of the particle inside the cluster, from now on
we shall visualize only the cluster part of the wave packet.

\begin{figure}[h]
$\begin{array}{cc}
{\resizebox{3.2in}{2.5in}{\includegraphics{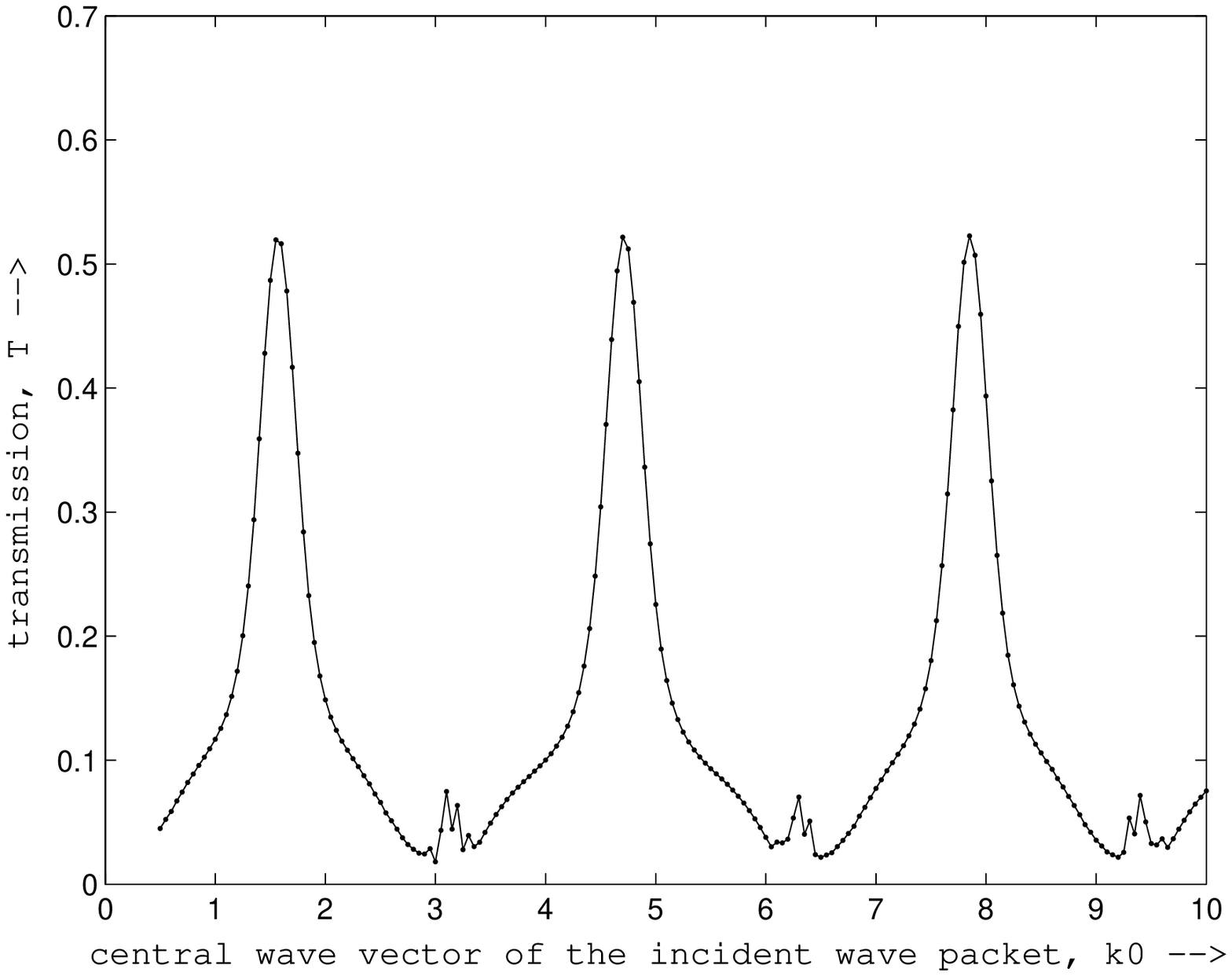}}} &
{\resizebox{3.2in}{2.5in}{\includegraphics{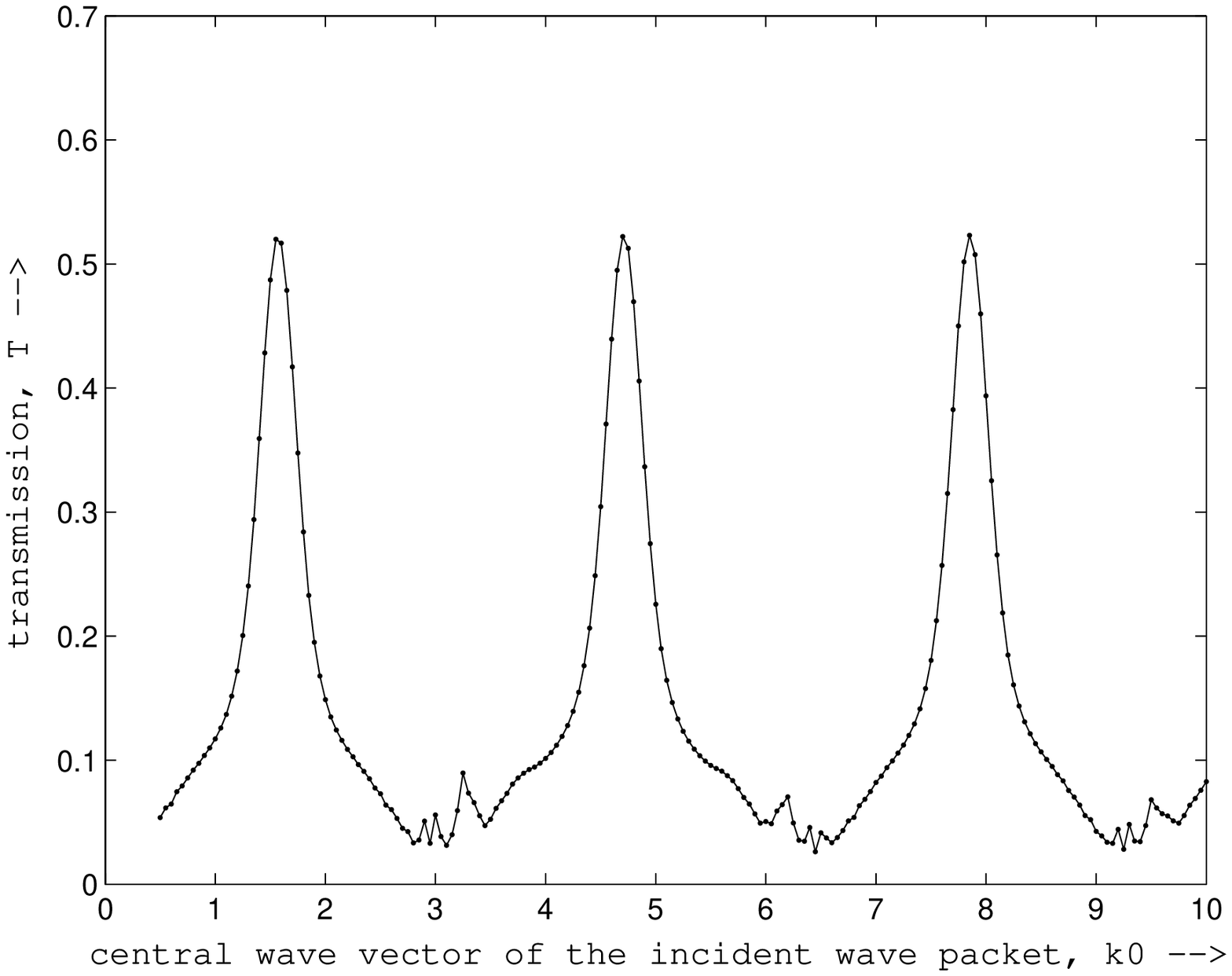}}} \\
(a) & (b)
\end{array}$
\caption{The effect of the central wave vector of the wave packet on transmission
         through cluster of size $70 \times 70$ connected to leads of length (a) 80
         and (b) 150.}
\label{sq_tvsk20}
\end{figure}

To study the effect of energy of the incident particle on transmission, we have
calculated the transmission coefficient for a range of central wave vectors of the
particle. Since we are interested in the behavior of the cluster in thermodynamic
limit, we also have investigated how the size of the cluster as well as the leads
affect transmission. We have studied three different sizes of clusters, namely
$20 \times 20$, $30 \times 30$ and $70 \times 70$ with four different
sizes of the leads, 50, 100, 150 and 200 connected diagonally with the clusters.
For each cluster-lead combination we have calculated transmission, T, by varying the
central wave vector from 0.5 to 10 (in units of inverse lattice constant) in increments
of 0.05. The width, $s$, of the incident packet in all these cases is taken to be equal
to 4 lattice constants. Since the largest size of the clusters considered in this work
is $70 \times 70$, we present in Fig.~\ref{sq_tvsk20} the results only for $70 \times 70$ cluster.

The above study suggests that the length of the leads does not have a major effect
on propagation beyond some finite length. For certain values of the central wave vector,
$k_{0}$, particle shows resonance transmission and $T$ reaches its maximum value. It is
also evident that a quantum particle may show zero transmittance even when the cluster
is completely ordered, a characteristic commonly known as resonance reflection. These
resonances are similar to the ones found by Cuansing and Nakanishi. \cite{cuansing:04}

\begin{figure}[htbp]
$\begin{array}{ccc}
{\resizebox{3.2in}{2.5in}{\includegraphics{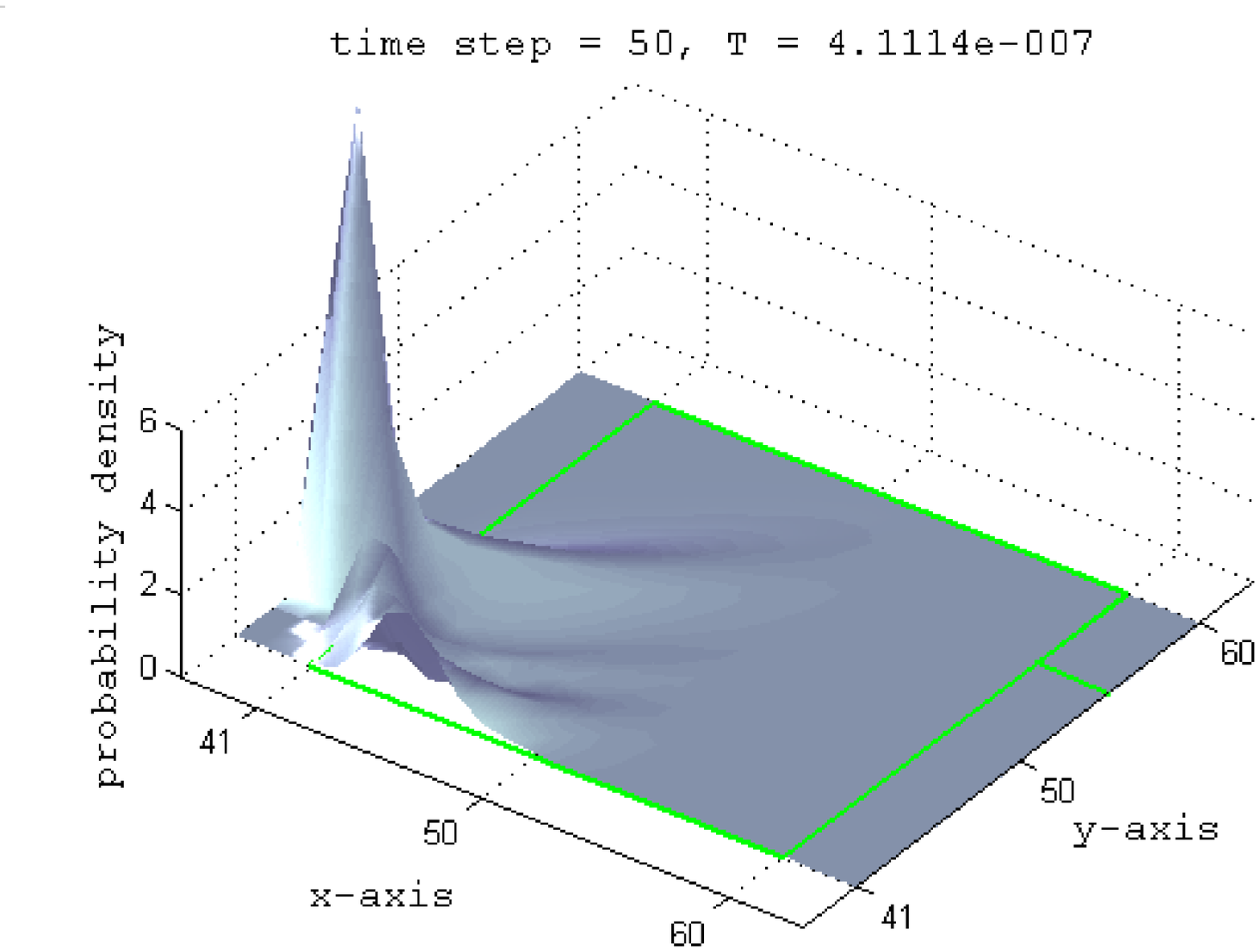}}} &
{\resizebox{3.2in}{2.5in}{\includegraphics{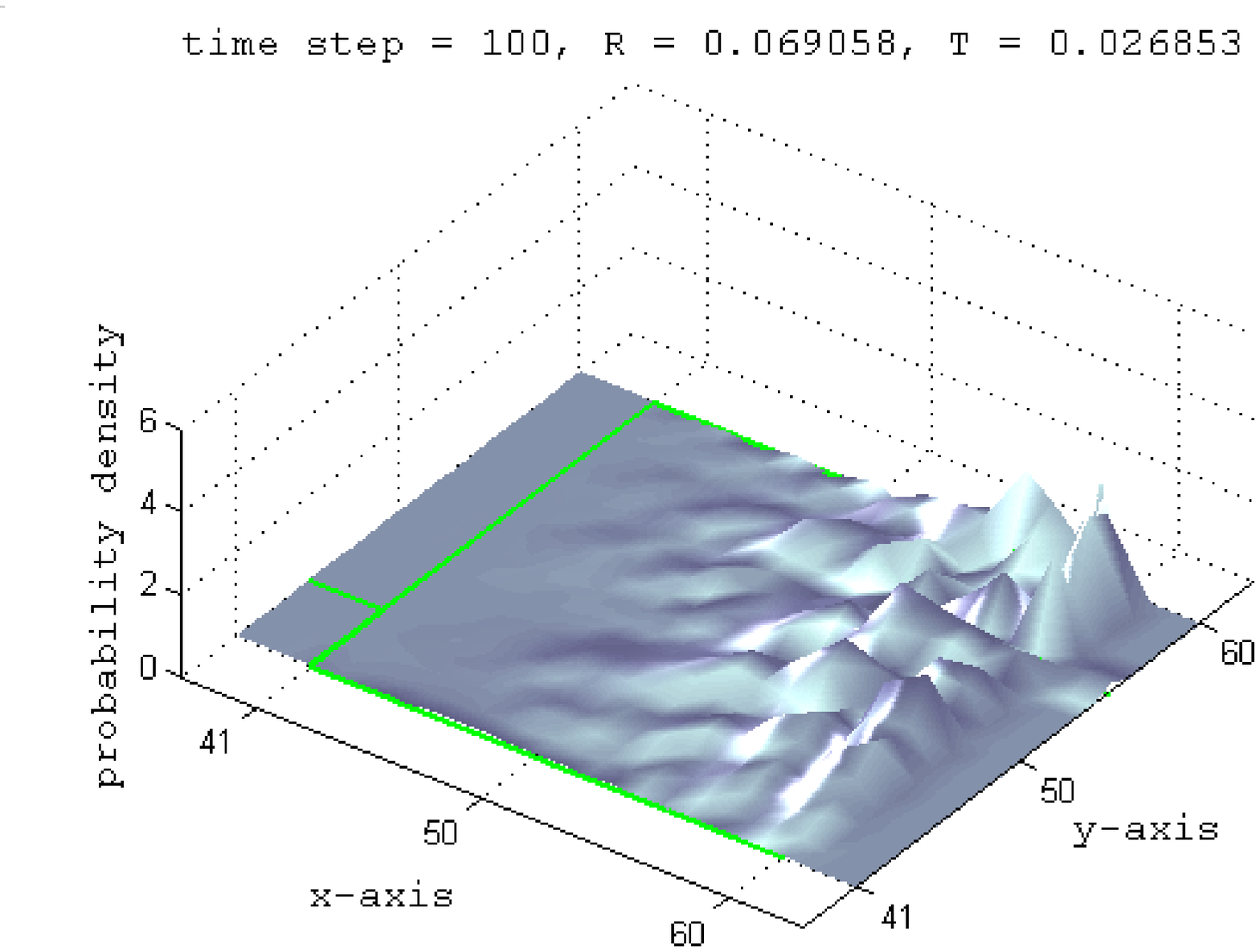}}}
\end{array}$
\caption{The time evolution of a wave packet through square cluster of size
         $20 \times 20$ with off diagonally connected leads.}
\label{p2p_offdiag}
\end{figure}

When the leads are connected to edge sites other than the diagonal corners,
transmission is generally reduced significantly. Fig.~\ref{p2p_offdiag} shows
the time evolution of the same wave packet as in Fig.~\ref{p2p_diag} when the
input lead is connected to the $5^{th}$ lattice point (input side) and the output
lead is at $15^{th}$ lattice point (output side) of the cluster. The reason for
this decrease in transmission is obvious from Fig.~\ref{p2p_offdiag}. As the
wave packet enters the cluster, it now spreads broadly throughout the whole cluster
and consequently only a small part of it reaches the output lead.

\begin{figure}[htbp]
{\resizebox{3.5in}{2.5in}{\includegraphics{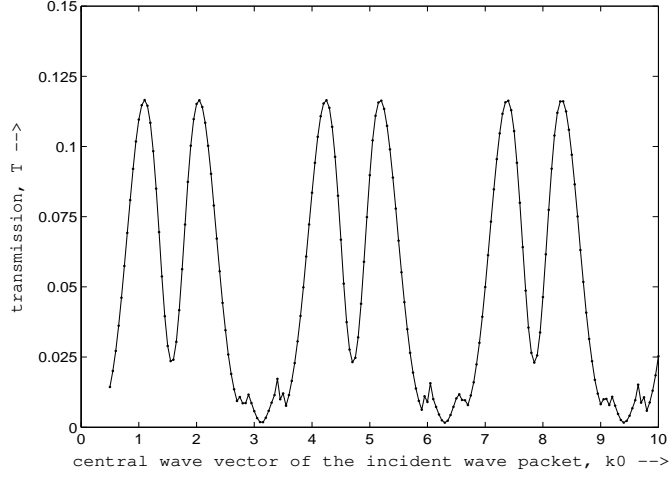}}}
\caption{The effect of the central wave vector of the wave packet on transmission
         through cluster of size $70 \times 70$ with input lead at $5^{th}$ lattice
         point on the input side and output lead at $65^{th}$ lattice point on the
         output side of the cluster.}
\label{offdiag_tvsk}
\end{figure}

The wave vector dependence of transmission for off-diagonally
connected leads is shown in Fig.~\ref{offdiag_tvsk}. The dependence
is quite different from what we have observed in the case of the
diagonal connection. The peak values of $T$ are significantly
reduced and the locations of the peaks are shifted as well, though
the same periodicity still exists as it must.

\begin{figure}[htbp]
$\begin{array}{cc}
{\resizebox{3.2in}{2in}{\includegraphics{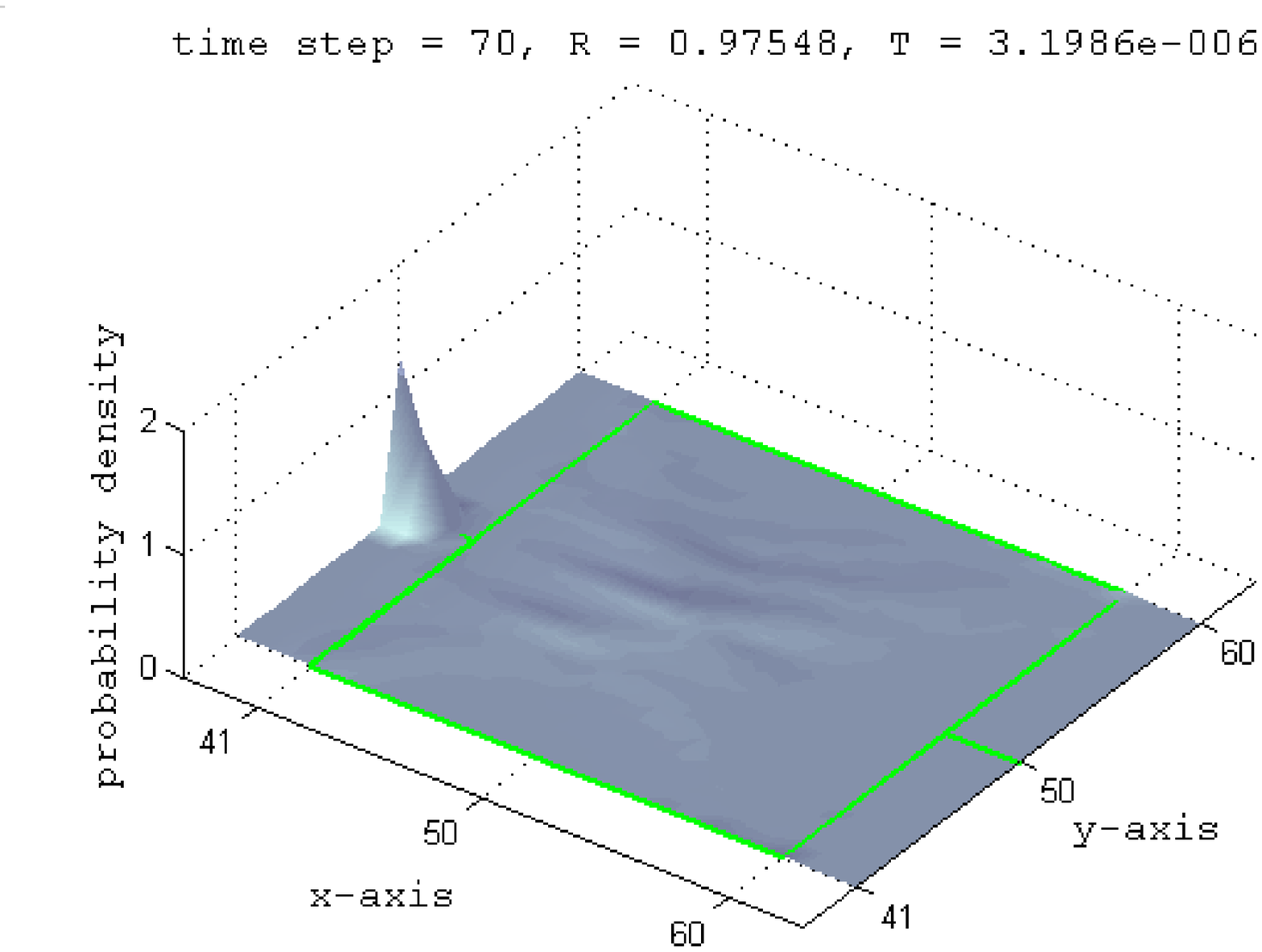}}} &
{\resizebox{3.2in}{2in}{\includegraphics{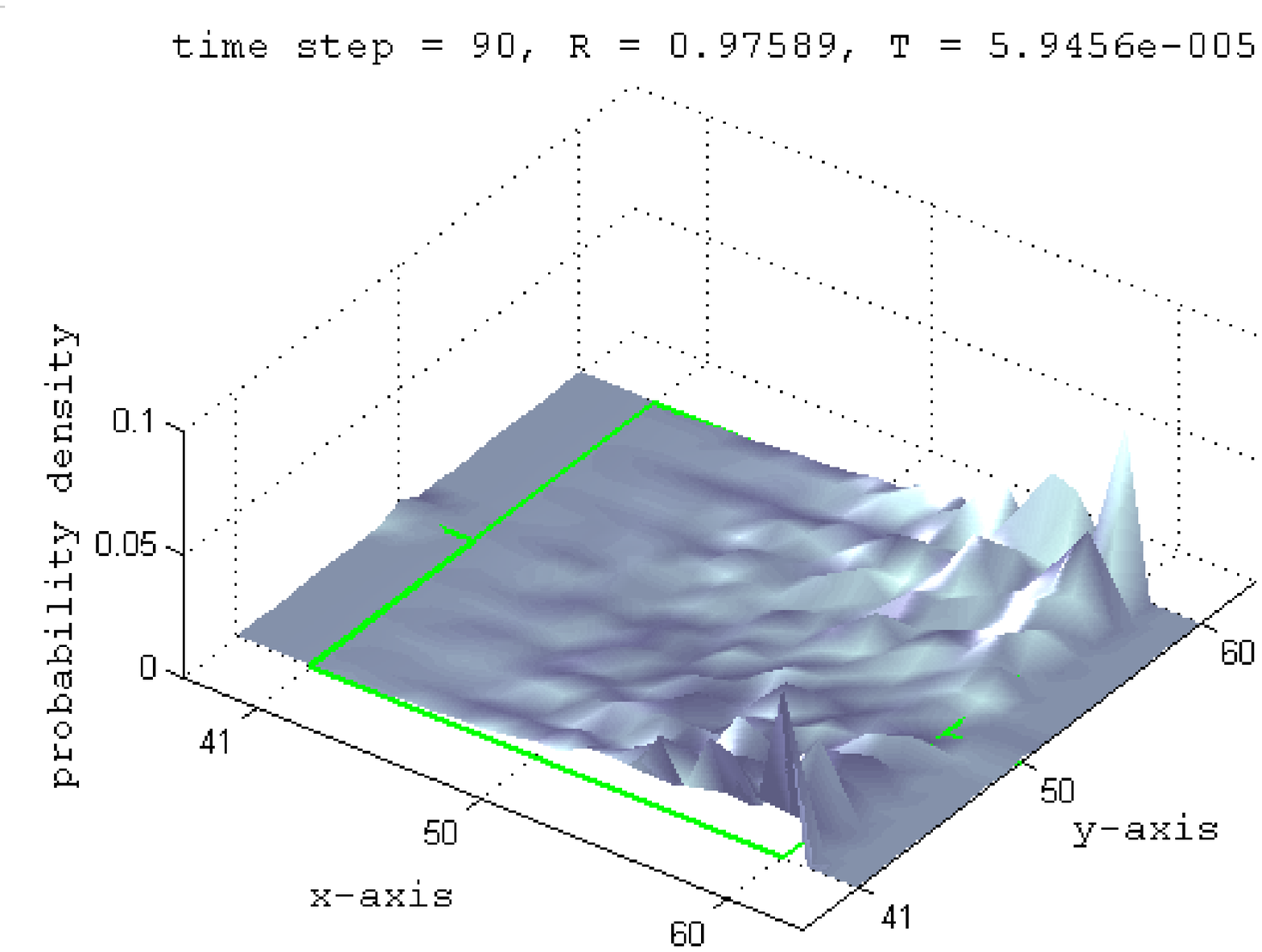}}} \\
\end{array}$
\caption{Transmission under busbar connection through a $20 \times
         20$ square cluster.}
\label{sq_busbar}
\end{figure}

The propagation of the wave packet under busbar type connection, on the other hand,
is dominated by the destructive interference of the packet on the edge of the cluster
at the input contact and very little transmission is observed (Fig.~\ref{sq_busbar}).
As before we display a packet centered about $k_{0}$ = 4.3 (in units of inverse
lattice constant) with cluster size $20 \times 20$. For certain energies, however,
transmission is significantly large (resonance transmission).

\subsection{Triangular Lattice}

\begin{figure}[htbp]
{\resizebox{3.5in}{2.2in}{\includegraphics{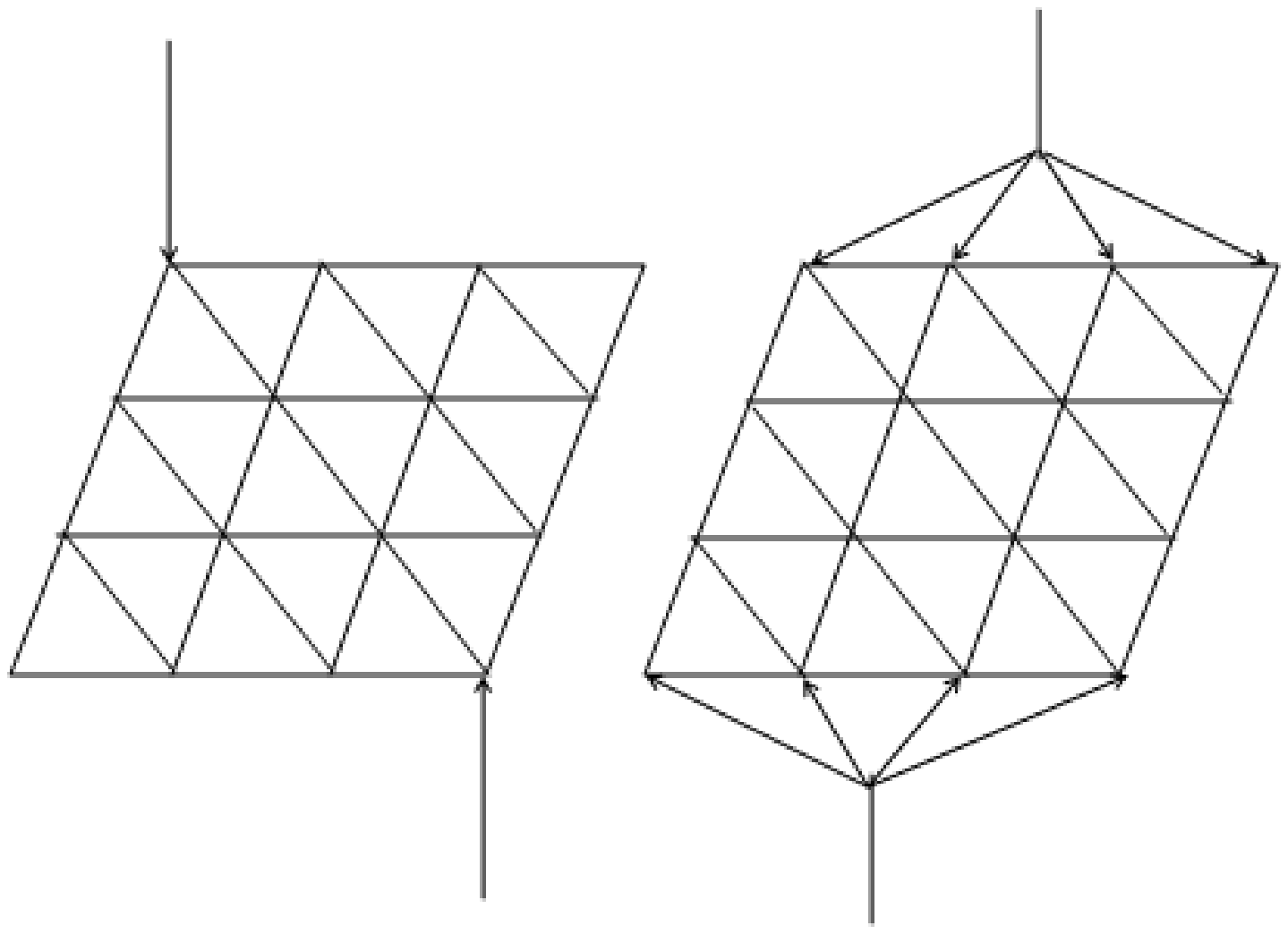}}}\\
(a)\hspace{1.5in} (b)
\caption{$4 \times 4$ Triangular Lattice: (a) point to point connection and
         (b) busbar type connection.}
\label{tri_lat}
\end{figure}

Though most of the results obtained in this work is based on square lattice, we have
also investigated the propagation of the particle on triangular lattice. The
triangular lattice (Fig.~\ref{tri_lat}) differs from the square lattice in that it
has 6 nearest neighbors for each lattice point and thus much higher local
connectivity. It is also not bipartite and thus may avoid features that are only
present on bipartite structures.\cite{inui94}

\begin{figure}[htbp]
$\begin{array}{cc}
{\resizebox{3.2in}{2.5in}{\includegraphics{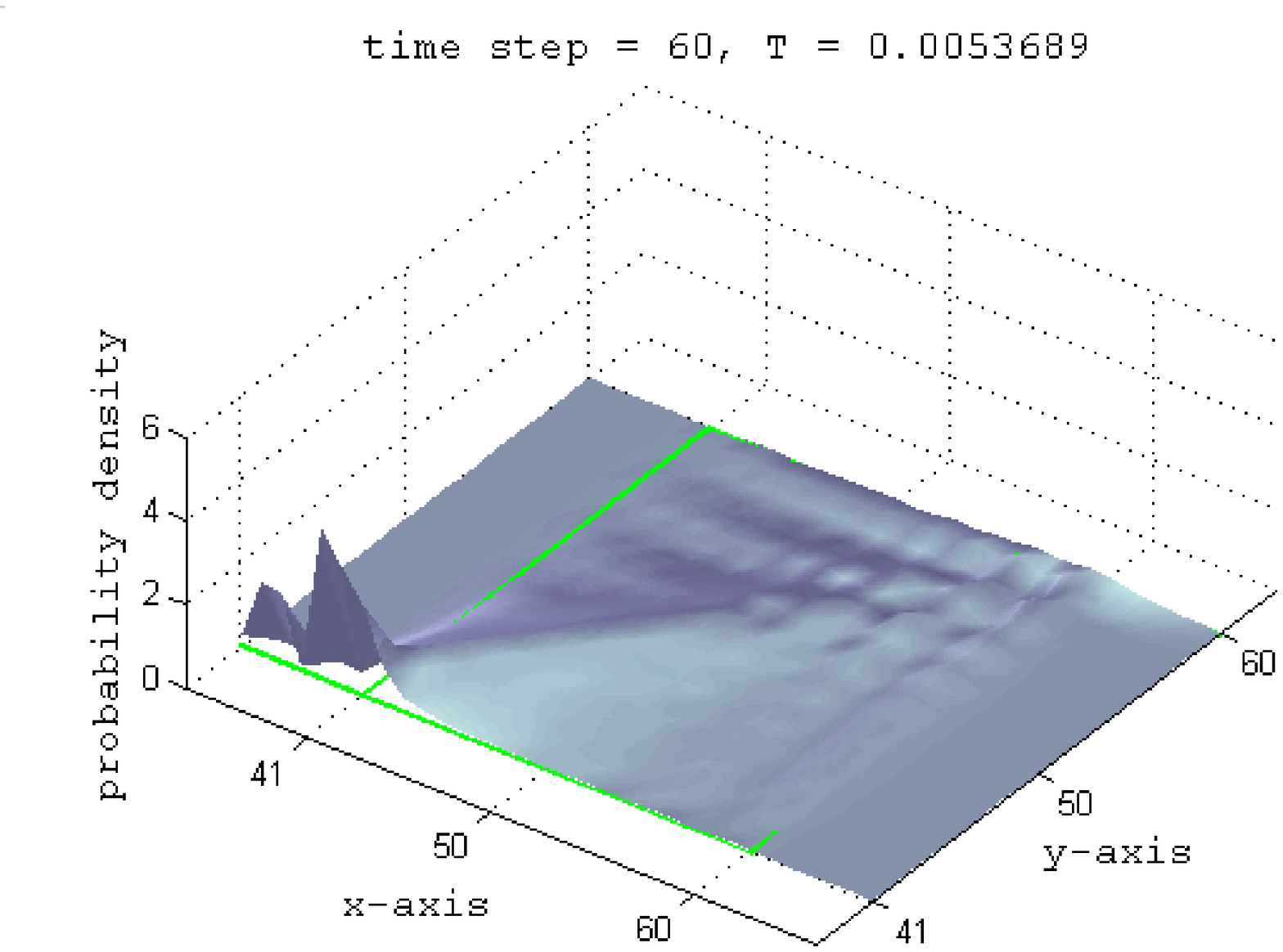}}} &
{\resizebox{3.2in}{2.5in}{\includegraphics{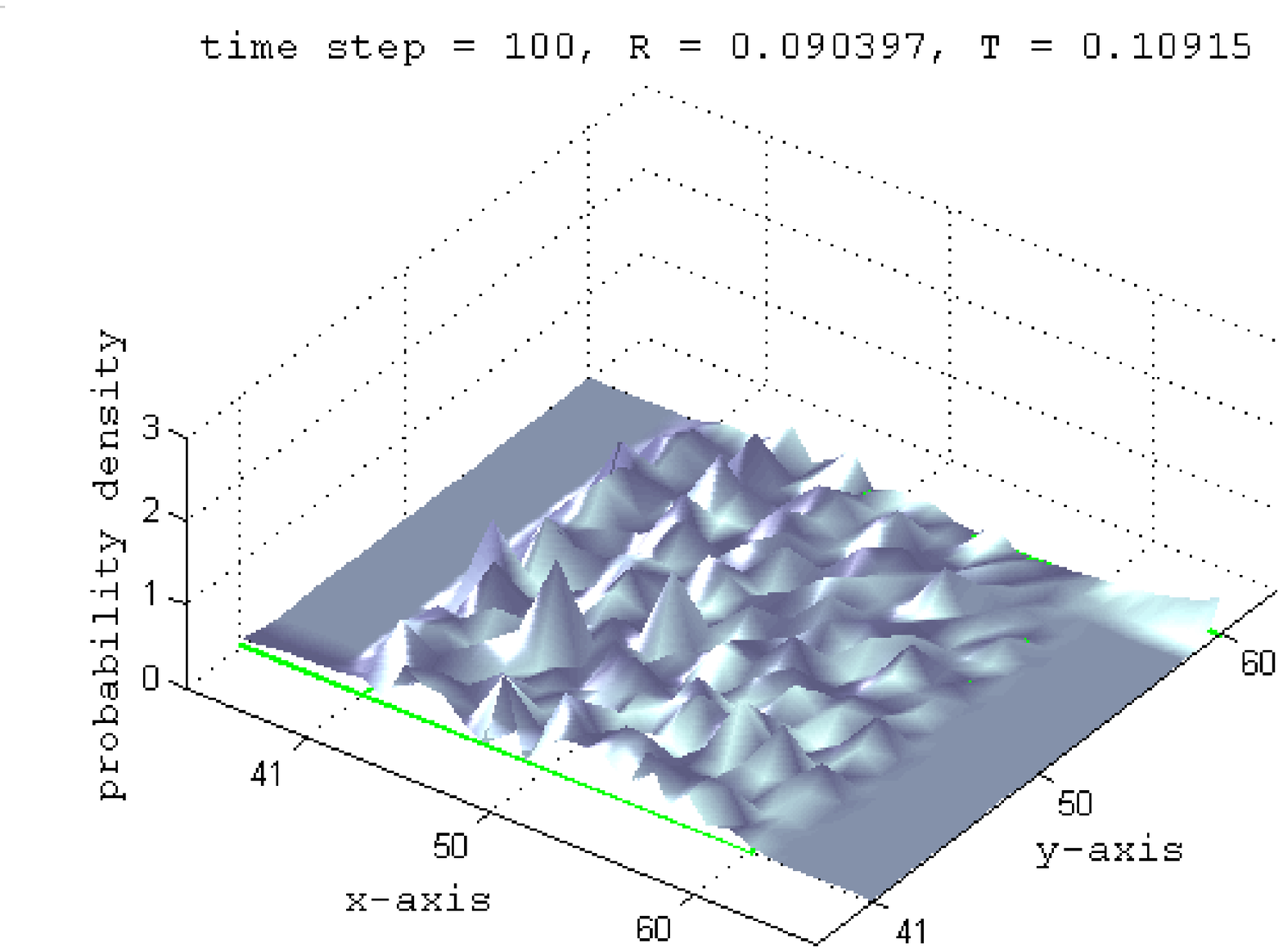}}} \\
\end{array}$
\caption{The time evolution of the wave packet through $20 \times 20$ triangular
         lattice with diagonally connected leads.}
\label{p2p_tri}
\end{figure}

While there are many more paths available for a particle on triangular lattice
than on square lattice, this may not necessarily lead to higher transmission unlike
in classical transport since the interference effect may is also expected to be larger.
We repeated our simulation shown in Fig.~\ref{p2p_diag} for the triangular lattice.
As is evident from the sample time evolution shown in Fig.~\ref{p2p_tri} with
diagonally connected leads, the spread of the wave packet throughout the cluster is
aided by the availability of additional paths. We found that these effects
generally result in lower transmission for the diagonal connection. However, transmission
increases with the busbar-type connection, which makes the difference in transmission
between these two connection types to be much smaller than for the square lattice.

\section{The effect of disorder on transmission}

\begin{figure}[htbp]
$\begin{array}{ccc}
{\resizebox{3in}{2.5in}{\includegraphics{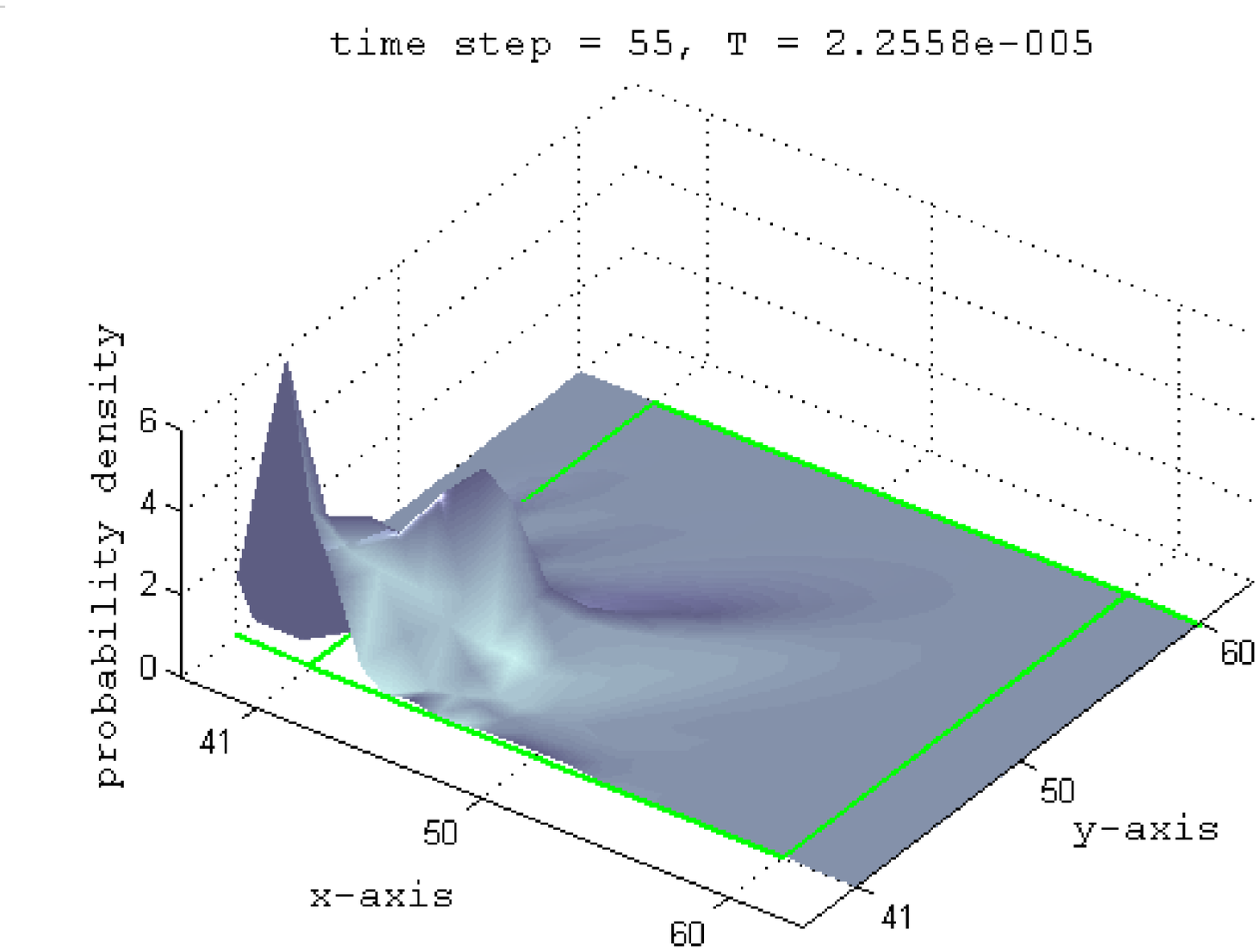}}} &
{\resizebox{3in}{2.5in}{\includegraphics{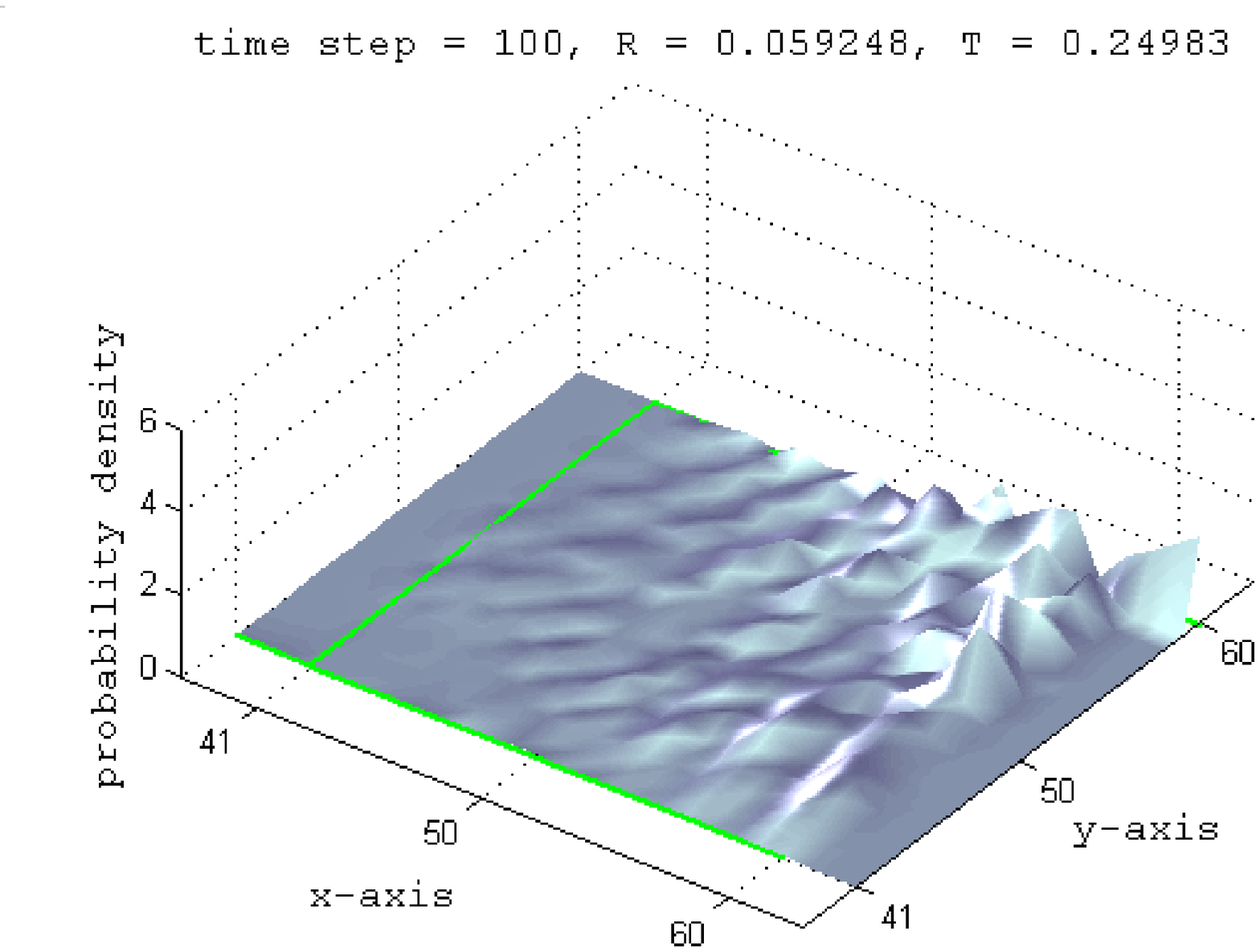}}}\\
\end{array}$
\caption{Time evolution of a wave packet propagation is shown in the presence of a
         single diluted site at the 86th lattice site near the body diagonal. The
         input and output leads are connected diagonally and this figure should be
         compared with Fig.~\ref{p2p_diag} for the corresponding time evolution
         without any dilution.}
\label{dilut1}
\end{figure}

Everything we have described so far was for completely ordered system. The main
reason why we have started this work is to visualize the effect of disorder on
transmission. As a first step towards understanding that, we investigate how
propagation is affected by prefixed dilution. A site is diluted by removing it from
the cluster, thus making it unavailable for hopping to. The diluted sites act as
infinite potential barriers for the particle. We start with a single diluted site
in a $20 \times 20$ cluster and observe how the wave packet with central wave vector
$k_{0}$ = 4.3 (in units of inverse lattice constant) evolves with time. The
Fig.~\ref{dilut1} shows some snapshots of the wave packet with a diluted site
located at $86^{th}$ lattice site which is close to the body diagonal of the cluster.

As the wave packet encounters the diluted site, it gets scattered by the infinite
barrier and consequently transmission is reduced. We investigated the effect for
other diluted sites and our study suggests that transmission is most strongly
affected if the diluted sites are close to the body diagonal of the cluster if the
leads are connected diagonally with the cluster. Such effects are illustrated
in Fig.~\ref{dilut1} which should be compared with Fig.~\ref{p2p_diag} for
the corresponding time evolution without dilution. The much broader spreading of
the wave packet is clearly observed in Fig.~\ref{dilut1} as well as in the
corresponding values of $T$.

\begin{figure}[htbp]
{\resizebox{3.5in}{2.2in}{\includegraphics{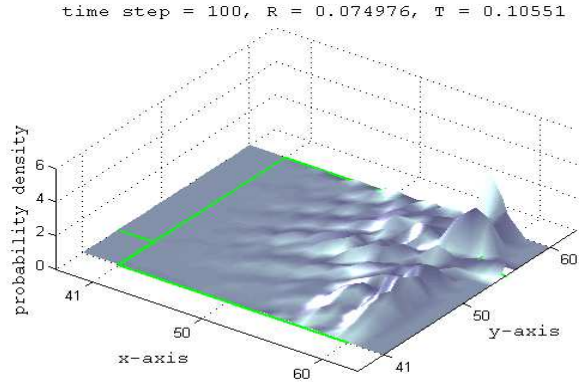}}}\\
\caption{Propagation in the presence of 3 diluted sites for an off-diagonal connection
         of the leads, where the input lead is connected to the $5^{th}$ lattice point
         on the input side of the cluster and the output lead is connected to
         $15^{th}$ lattice point on the output side of the cluster.}
\label{dilut2}
\end{figure}

While the effect of dilution is generally to reduce transmission through the cluster
for diagonally connected leads, it may enhance transmission at least for some
off-diagonally connected leads. If we repeat the simulation shown in
Fig.~\ref{p2p_offdiag} with 3 diluted sites at $27^{th}$, $376^{th}$ and
$396^{th}$ lattice points, the propagation appears as shown in Fig.~\ref{dilut2}
(Since sites are labeled with numbers as illustrated in Fig.~\ref{sq_lat}, these
sites are off-diagonal and near either point of contact with the input and output
chains). Though the qualitative pictures of the extent of the wave packet spreading
appear similar in Fig.~\ref{p2p_offdiag} (without dilution) and Fig.~\ref{dilut2}
(with 3 sites diluted), the corresponding values of $T$ at the 100th time step show
that transmission is actually enhanced in the presence of these diluted sites.

\begin{figure}[htbp]
$\begin{array}{cc}
{\resizebox{3.2in}{2.2in}{\includegraphics{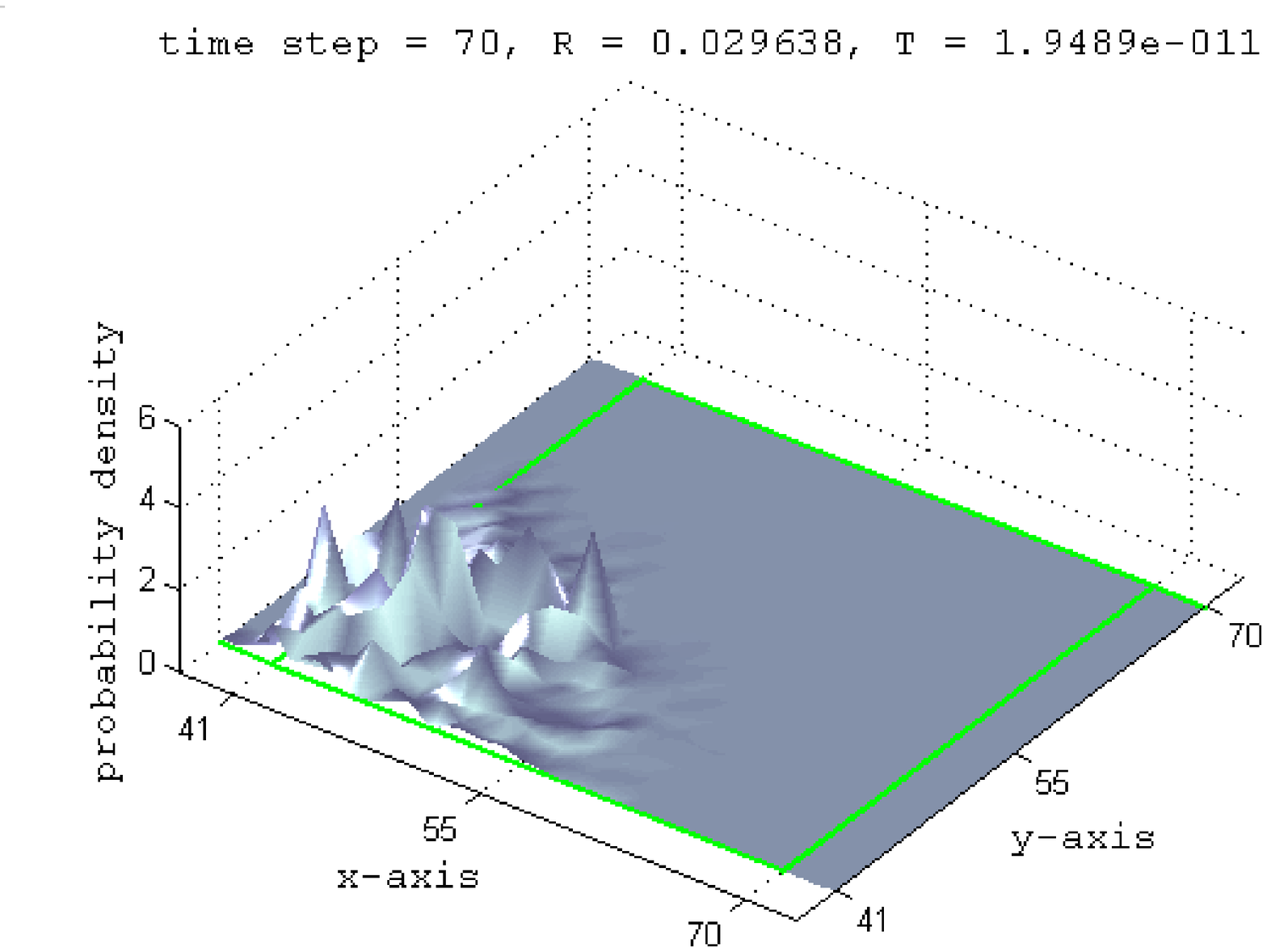}}} &
{\resizebox{3.2in}{2.2in}{\includegraphics{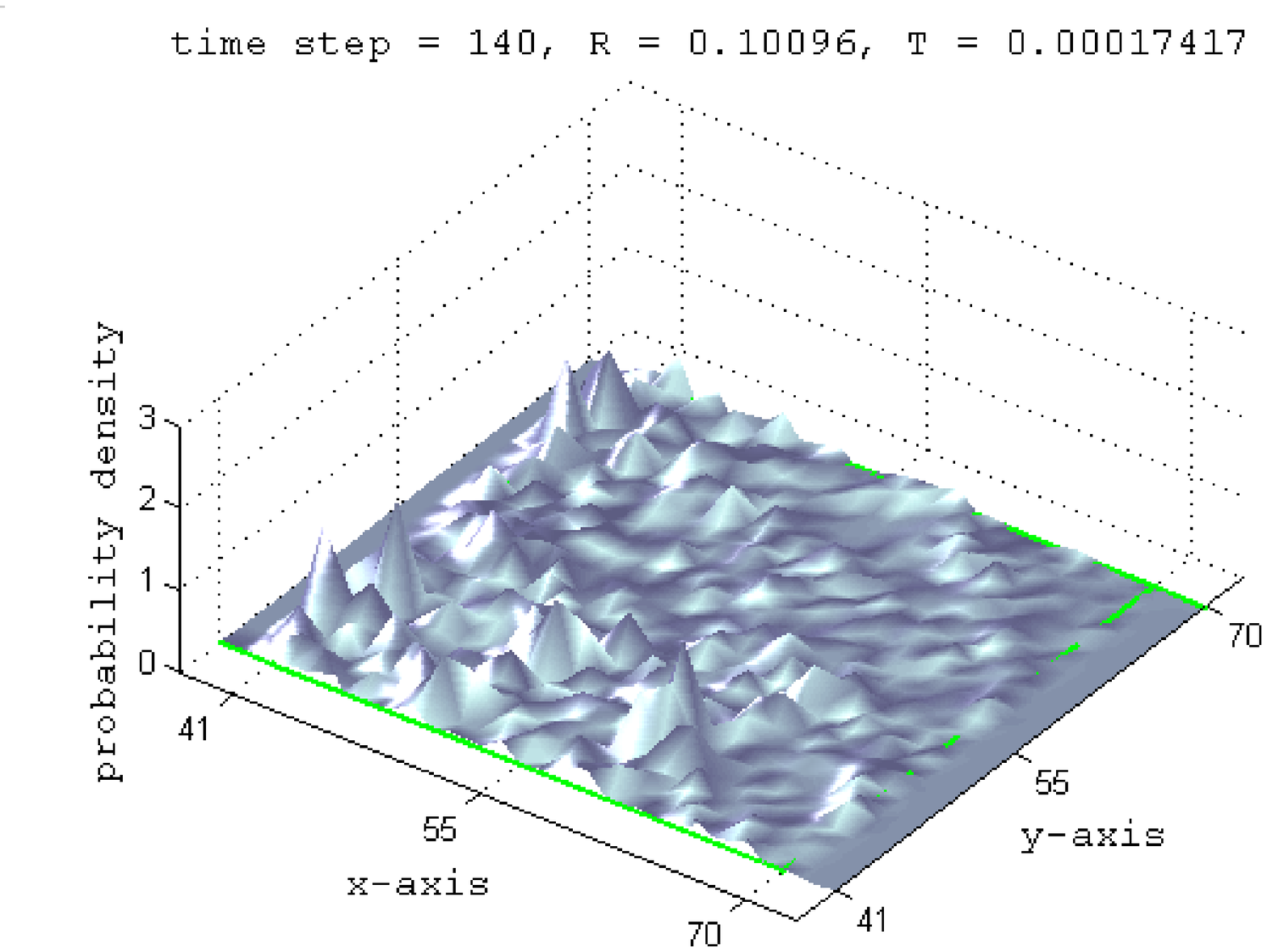}}}\\
\hspace{3in}(a)\\
{\resizebox{3.2in}{2.2in}{\includegraphics{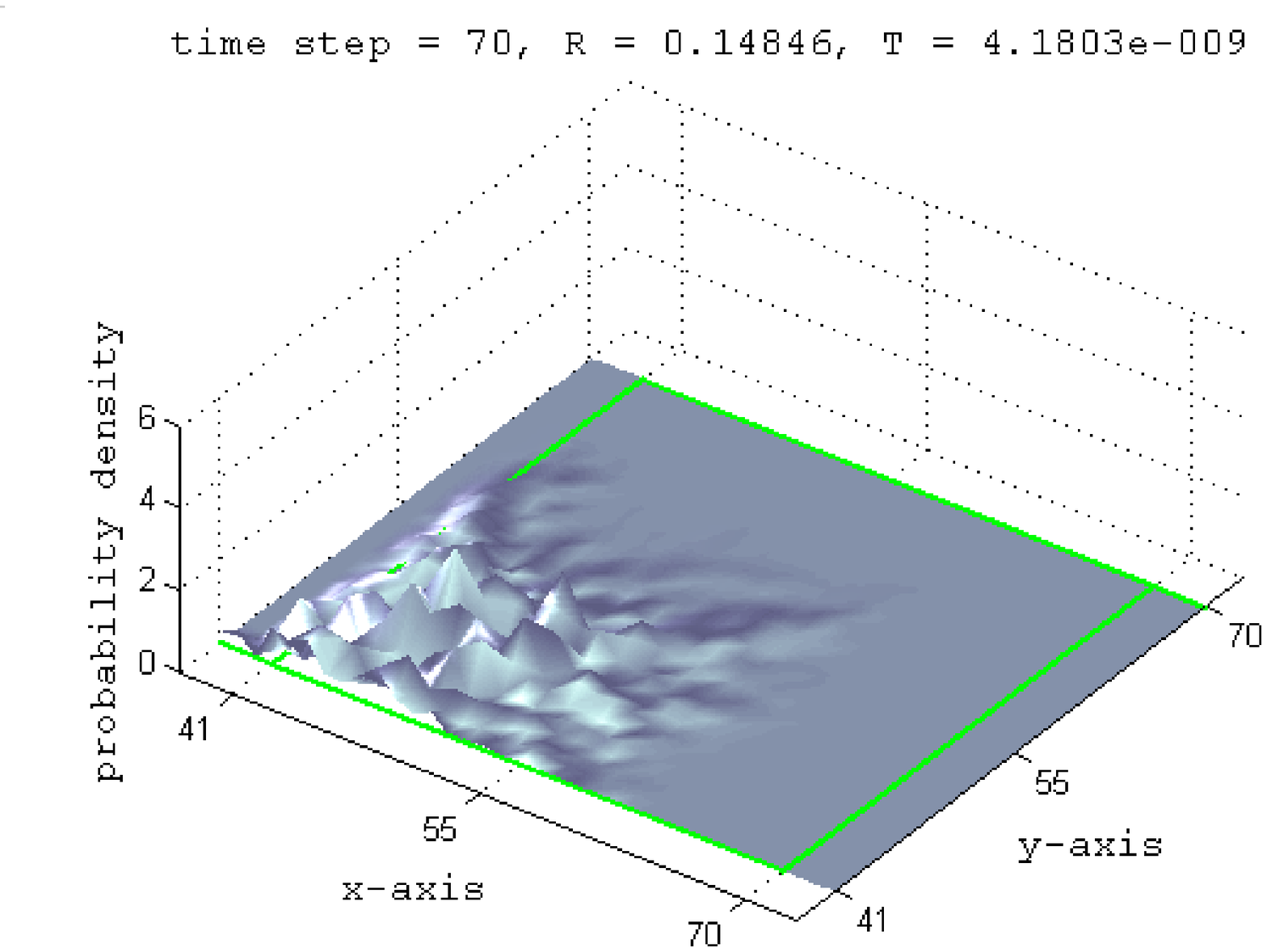}}} &
{\resizebox{3.2in}{2.2in}{\includegraphics{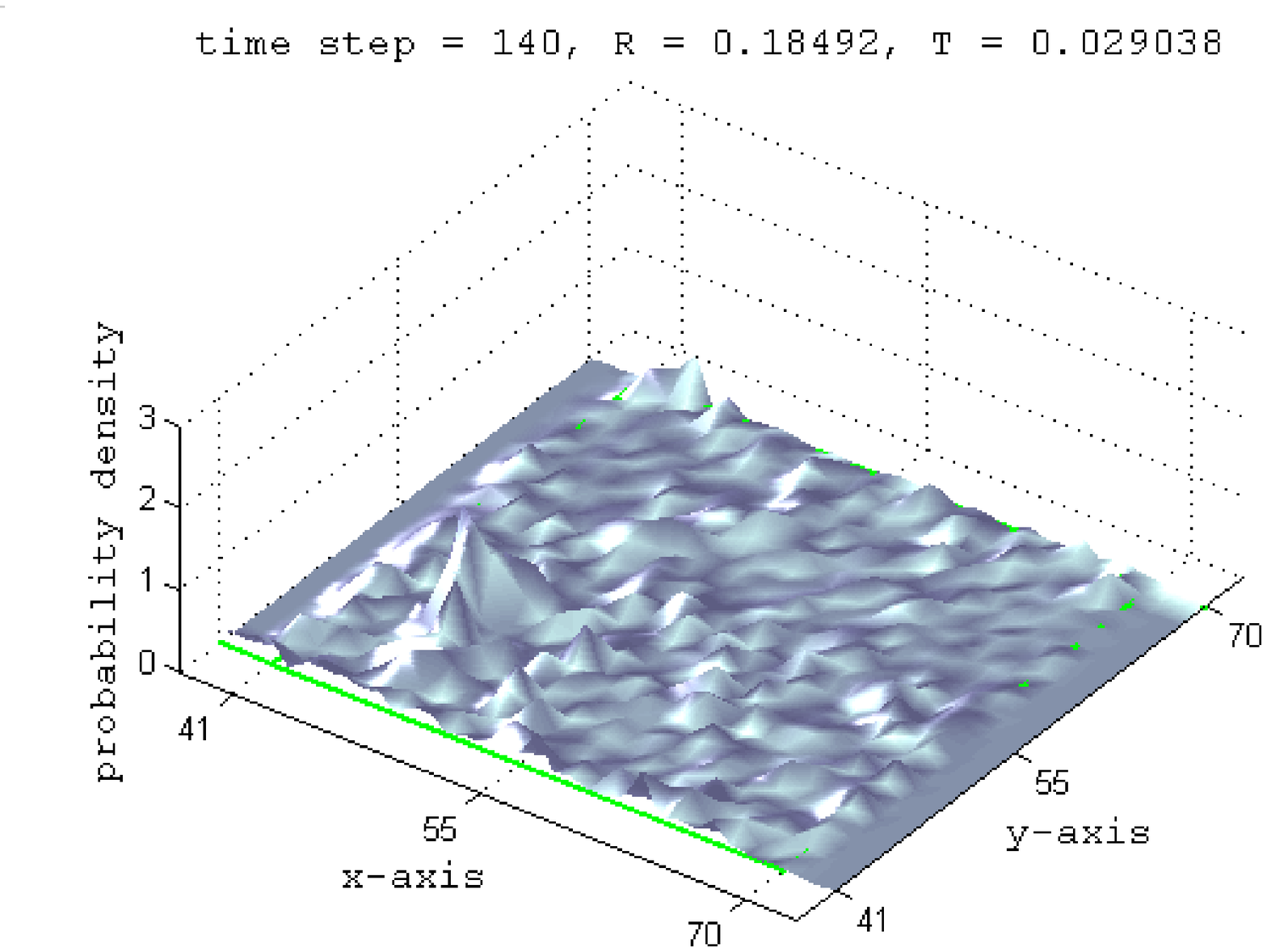}}}\\
\hspace{3in}(b)\\
\end{array}$
\caption{Propagation of a wave packet through a $30 \times 30$ cluster with $10\%$
         disorder with diagonally connected input and output leads. (a) and (b) are
         two independent realizations. Note that the values of $T$ vary a great deal
         in the two realizations although there are similarities in the qualitative
         features of the wave packet spreading. }
\label{disorder}
\end{figure}

The study of propagation under the dilution of specific sites gives us some clue on
how disorder may affect propagation through the cluster. We introduce disorder in
our system by removing sites from the cluster randomly with a given probability.
For instance, a $10\%$ disorder means a site is removed if a random number,
uniformly generated between 0 and 1, associated with that site is less than 0.1.
The Fig.~\ref{disorder} shows two independent realizations of propagation through
a cluster of size $30 \times 30$ with $10\%$ disorder with diagonally connected
leads. As noted in the figure, the values of transmission $T$ varies widely from
realization to realization. If the concentration of diluted sites near the input
lead is large then the reflection coefficient is observed to be large. Another
important observation is that a large part of the wave packet is trapped inside
the disordered cluster.

\begin{figure}[htbp]
{\resizebox{3.8in}{2.8in}{\includegraphics{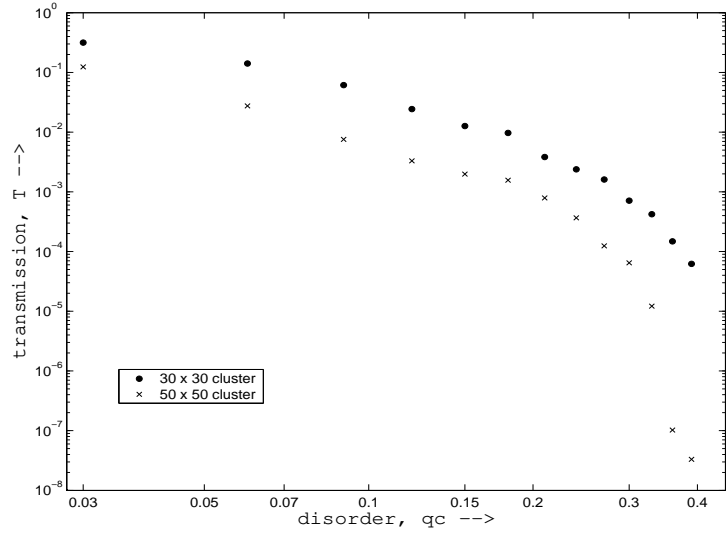}}}\\
\caption{Dependence of transmission $T$ on disorder is illustrated for clusters of size
         $30 \times 30$ and $50 \times 50$. Every data point is the average over 100
         independent realizations. The apparent onset of a sharp fall in $T$ as disorder
         increases motivates our closer look as given in the remainder of this work.}
\label{tvsq}
\end{figure}

This method can be used to study the presence or absence of a localized-delocalized
transition in 2D disordered quantum percolation system. As a first step we have
applied this method to study the behavior of a particle under different amount of
disorder for a given cluster. For most of the work that follows we have used diagonally
connected leads of length 80 and a wave packet of central wave vector $k_{0}$ = 4.7
(in units of inverse lattice constant) with width $s$ = 4 lattice constants. To
minimize the effect of the boundary on the interior property of the disordered
clusters, we made good contacts by keeping the nine sites nearest to both the input
and the output contact points always occupied (and available). We considered two
different sizes of clusters, $30 \times 30$ and $50 \times 50$ and for each size
and each dilution, 100 independent realizations are used to obtain average values
of $T$. The results are plotted in Fig.~\ref{tvsq}.

We observe from Fig.~\ref{tvsq} that the rate of decrease in transmission as disorder
is increased is small if disorder is below about $20\%$ but, as the disorder
increases beyond $20\%$, transmission begins to fall sharply. This suggests that
even if there is no localized-delocalized transition in 2D, the nature of
localization of the particle in the disordered cluster may not be the same at all
strengths of disorder.

\begin{figure}[h]
{\resizebox{5.25in}{4.1in}{\includegraphics{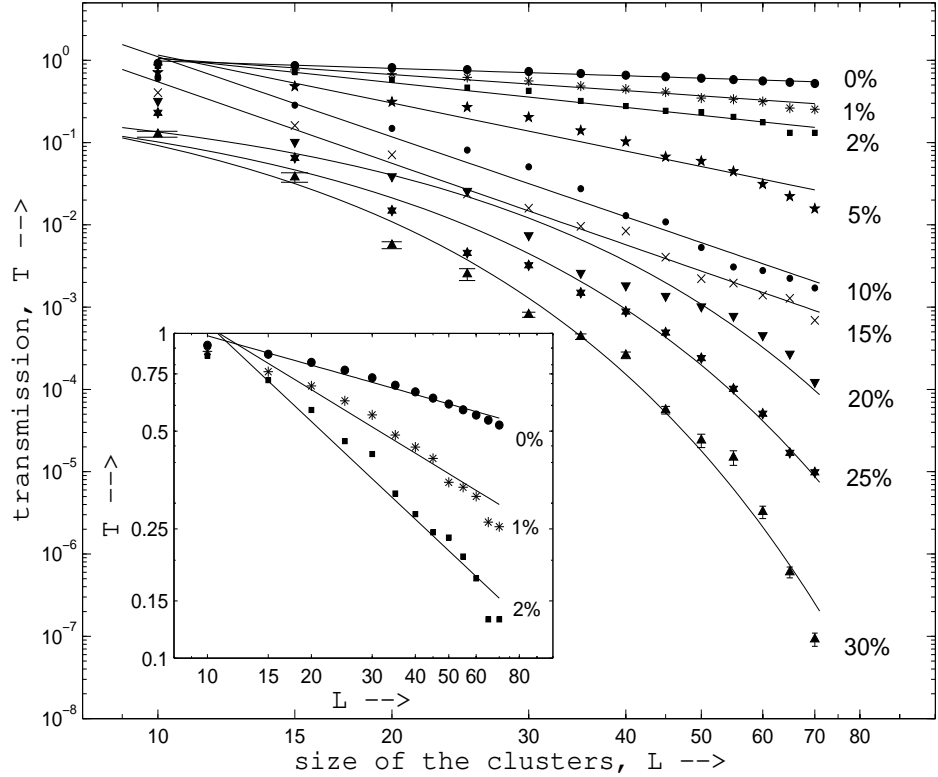}}}
\caption{The finite-size scaling of transmission through disordered clusters with
         diagonally connected leads. In the main panel a log-log plot of $T$ vs $L$
         is shown for various amounts of disorder. The solid lines are from the fits
         made on the log-log scale resulting in the parameters summarized in
         Table~\ref{tab1}, and the standard errors of the mean are smaller than the
         size of the symbols except where explicitly plotted. The inset shows the
         the transmission curves only for low disorders (up to 2\%) with an expanded
         scale in the vertical direction. Every data point is the average over 100 realizations.}
\label{scal1a}
\end{figure}

To study the presence or absence of a delocalization transition, one needs to
investigate the behavior of the 2D system in thermodynamic limit. This, however,
is not possible in numerical methods, and therefore, we resort to a finite size
scaling approach in which we calculate the transmission while gradually increasing
the size of the system for a given amount of disorder. The result is then
extrapolated to study the bulk behavior in the thermodynamic limit. We first obtain
the transmission curves for many levels of disorder with the leads connected
diagonally with the cluster. The scaling results for this simulation are plotted
in Fig.~\ref{scal1a} where each data point is the average of 100 disorder realizations.
The errors associated with most points are smaller than the symbols used except
for those few cases where they are explicitly shown in the plot. In this work,
the maximum size of the cluster that could be reached was $70 \times 70$.

We have attempted to fit the results to power laws and exponentials in all cases
and found that either one of these functions fits clearly better than the
other in different regions of the disorder. The resulting best fits are plotted
in Fig.~\ref{scal1a}, and the best fitting parameters are summarized in
Table~\ref{tab1} together with the fitting ranges for the $95\%$ confidence bounds
in parentheses.

\begin{table}[h]
\caption{Table for fitting parameters performed on the log-log scale for diagonally connected
leads without allowing for a possibility of a non-zero offset. Shown in the parentheses are
the lower and upper bounds for 95\% confidence level.}
\label{tab1}
\begin{center}
\begin{tabular}{||c|c|c|c|c||} \hline
Disorder & Fit        & \multicolumn{2}{c|}{Parameters}        &$|R|^{2}$ \\   \cline{3-4}
         & equation   & a                 & b                  &          \\ \hline
$0\%$    &            &1.96 (1.72, 2.23)  &0.30 (0.27, 0.34)  &0.97       \\ \cline{1-1} \cline{3-5}
$1\%$    &            &4.71 (3.26, 6.79)  &0.66 (0.56, 0.76)  &0.95       \\ \cline{1-1} \cline{3-5}
$2\%$    &$T = a\cdot L^{-b}$ &10.7 (6.55, 17.6)  &1.00 (0.86, 1.14)  &0.96  \\ \cline{1-1} \cline{3-5}
$5\%$    &            &101  (30, 338)     &1.94 (1.60, 2.28)  &0.94       \\ \cline{1-1} \cline{3-5}
$10\%$   &            &1883 (534, 6640)   &3.23 (2.88, 3.58)  &0.97       \\ \cline{1-1} \cline{3-5}
$15\%$   &            &1071 (486, 2359)   &3.29 (3.07, 3.51)  &0.99       \\ \hline
$20\%$   &            &0.45 (0.21, 0.95)  &0.12 (0.10, 0.14)  &0.96       \\ \cline{1-1} \cline{3-5}
$25\%$   &$T = a\cdot\mbox{e}^{-b L}$ &0.48 (0.27, 0.87)  &0.16 (0.14, 0.17)  &0.98  \\ \cline{1-1} \cline{3-5}
$30\%$   &            &0.78 (0.35, 1.72)  &0.21 (0.20, 0.23)  &0.98       \\ \hline
\end{tabular}
\end{center}
\end{table}

We observe that at lower disorders ($15\%$ and below) transmission
appears to fit well with straight lines in log-log plots (i.e., power laws
$T \sim a L^{-b}$), but as the disorder increases, the exponential fits
($T \sim a \mbox{e}^{-bL}$) become better, consistent with the states becoming
exponentially localized in the thermodynamic limit. In Table~\ref{tab1}, $|R|^2$ in
the last column is obtained from the relation $1 - |R|^2 \equiv \sigma_{y|x}^2/\sigma_y^2$
where $\sigma_{y|x}$ is the standard deviation of the data scatter about the best
fits and $\sigma_y$ is the standard deviation of the overall data. Thus $|R|^2$ is
a measure of the fraction of the data scatter that is explained by the best fit
functional form, and $R$ reduces to the usual correlation coefficient for the case
of linear regression. It should be noted that the fitting procedures summarized in
Table~\ref{tab1} were performed on the log-log scale and then the parameters so-obtained
were converted to the linear scale. We may note that some of the data points are
$2\sigma$ or more away from the fits for some values of disorder. Part of this is due
to the discrete nature of the system which leads to oscillatory behavior of $T$, but
part of this is also an indication that these fits were made without allowing for a
possible, non-zero offset as discussed further below.

To see the finer details of transmission curves in Fig.~\ref{scal1a} for lowest disorder amounts,
we have plotted them separately in the inset. We observe from the inset that the power-law fits
(straight lines in these plots) are not as good as they appear on the main part of the figure,
but the data from the lowest disorder amounts follow a very similar trend
to the zero-disorder case (also shown) where a non-zero transmission is expected.

\begin{figure}[h]
{\resizebox{5.25in}{4.1in}{\includegraphics{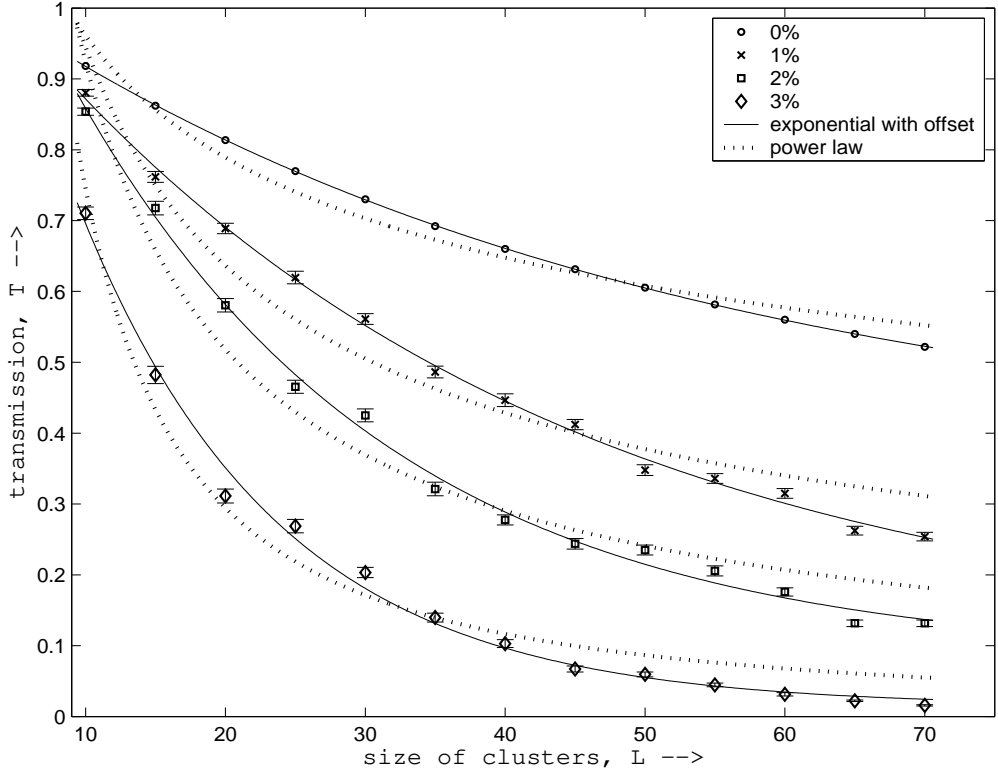}}}
\caption{The linear plot of the finite size scaling data of transmission through
         disordered clusters with diagonally connected leads for low disorder. The
         dotted lines represent the power-law fits and the solid lines represent
         exponential fits with finite offsets, both performed on the linear scale
         as direct, non-linear fits. For this range of disorder, an exponential with
         a finite offset seems to fit clearly better than a power-law, suggesting
         a non-zero transmission $T$ even in the thermodynamic limit.}
\label{scal1b}
\end{figure}

To further investigate the matter we have plotted the data on the linear scale
as shown in Fig.~\ref{scal1b}. Each data set is then fitted directly
(and non-linearly) to a power law (dotted lines) and an exponential with a possible offset
(solid lines) . The fit parameters  for these non-linear fits are summarized in Table~\ref{tab2}.
To compare the goodness of fits we need to consider not only the curves but the $|R|^{2}$
and SSE (sum square error) values as well, and it is evident from Table~\ref{tab2} that in
all these three counts an exponential with offsets gives significantly better fits than
a power law. Except for the 5\% dilution case, the 95\% confidence range of the offsets $c$
clearly exclude zero according to these fits, consistent with a non-zero $T$ even in
thermodynamic limit for these cases. Thus the lower disorder portion of the data that was fitted
to a power-law in Fig.~\ref{scal1a} in fact splits into those that actually fit better an
exponential with offset (extended - up to $\sim 5\%$ disorder) and those that fit a
power-law better (power-law localized - between about $\sim 5\%$ and $\sim 20\%$ disorder).

\begin{table}[h]
\caption{Fitting parameters of non-linear fits for diagonal connection at low disorder.
Shown in the parentheses are the lower and upper bounds for 95\% confidence level.}
\label{tab2}
\begin{center}
\begin{tabular}{||c|c|c|c|c|c|c|c|c|c||}   \hline
Disorder   & \multicolumn{4}{c|}{Power law fit: $T = a\cdot L^{-b}$}  & \multicolumn{5}{c||}{Exponential fit with offset: $T = a\cdot\mbox{e}^{-b L} + c$ } \\ \cline{2-10}
           & a               & b               & $|R|^{2}$& SSE  & a               & b                & c                 &$|R|^{2}$& SSE   \\ \hline
   0\%     &1.85 (1.64, 2.07)&0.28 (0.25, 0.32)& 0.96     &0.007 &0.68 (0.67, 0.68)&0.02  (0.02, 0.02)&0.36 (0.35, 0.37)  &1        & 0.000 \\ \hline
   1\%     &3.52 (2.56, 4.47)&0.57 (0.48, 0.66)& 0.95     &0.024 &1.01 (0.97, 1.05)&0.026 (0.02, 0.03)&0.09 (0.02, 0.16)  &0.99     & 0.001 \\ \hline
   2\%     &6.31 (4.16, 8.46)&0.83 (0.72, 0.95)& 0.96     &0.025 &1.20 (1.13, 1.28)&0.044 (0.03, 0.05)&0.08 (0.04, 0.12)  &0.99     & 0.002 \\ \hline
   5\%     &16.2 (8.47, 24)  &1.34 (1.16, 1.52)& 0.97     &0.015 &1.37 (1.22, 1.53)&0.070 (0.06, 0.08)&0.01 (-0.008, 0.03)&0.99     & 0.003 \\ \hline
\end{tabular}
\end{center}
\end{table}

\begin{figure}[h]
{\resizebox{5.25in}{4.1in}{\includegraphics{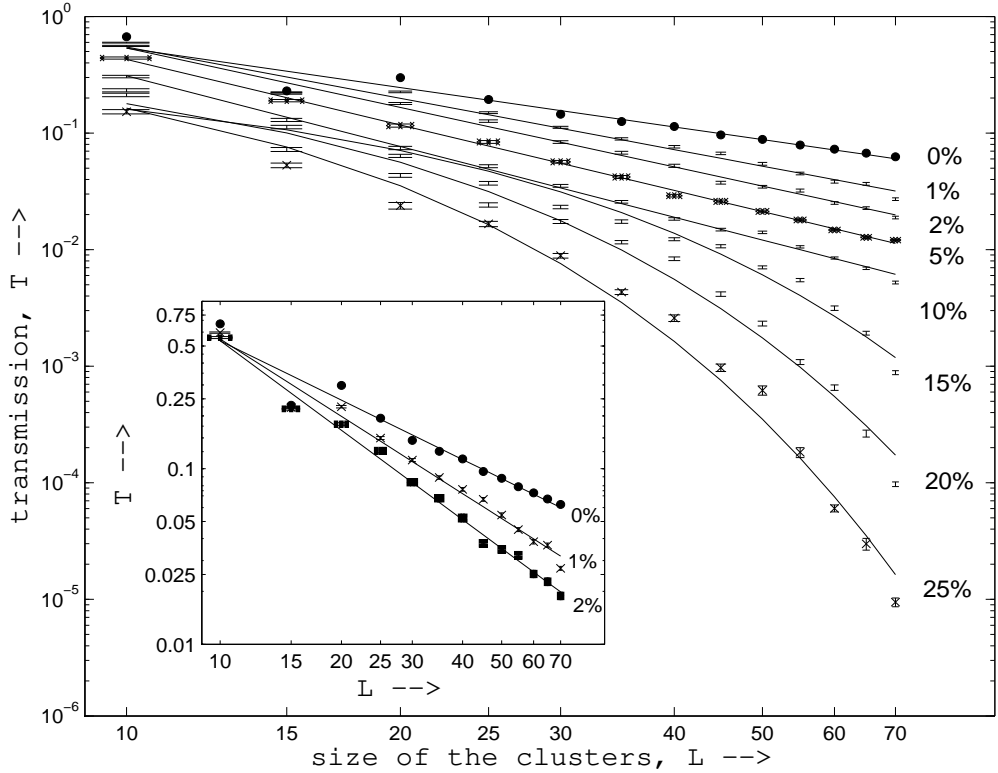}}}
\caption{The finite size scaling of transmission through disordered clusters with
         off-diagonally connected leads (the input lead connected at the $5^{th}$ lattice
         site on the input side and the output lead is at the $(L -5)^{th}$ lattice
         site on the output side of the cluster, where $L$ is the linear size of clusters).
         The main panel shows the log-log plot of $T$ vs $L$ for different disorder
         amounts, where the solid lines are the best fits performed on the log-log scale
         similarly to those of the diagonal connection in Fig.~\ref{scal1a}.
         The standard errors of the mean are smaller than the size of the symbols
         except where explicitly plotted. The inset shows the transmission
         curves only for the lowest disorders with an expanded vertical scale.
         Every data point is the average over 100 realizations.}
\label{scal2a}
\end{figure}

We have repeated the above procedure for an off-diagonal connection where
the input lead is connected at the $5^{th}$ lattice site on the input side and
the output lead at the $(L -5)^{th}$ lattice site on the output side ($L$ being
the linear size of the clusters). The results for different disorder amounts are
plotted in Fig.~\ref{scal2a}.
The central wave vector of the incident wave packet in this simulation is $k_{0}$ = 3.6,
keeping all other parameters unchanged. It is clear from the
Fig.~\ref{scal2a} that the scaling behavior is generally quite similar to that of the diagonal
connection; however, the inset appears to indicate that deviations from the power-law fits
(straight lines) for the lowest disorder amounts (up to 2\%) are much less than the case
of the diagonal connection shown in the inset of Fig.~\ref{scal1a} for similar values of disorder.

\begin{figure}[h]
{\resizebox{5.25in}{4.1in}{\includegraphics{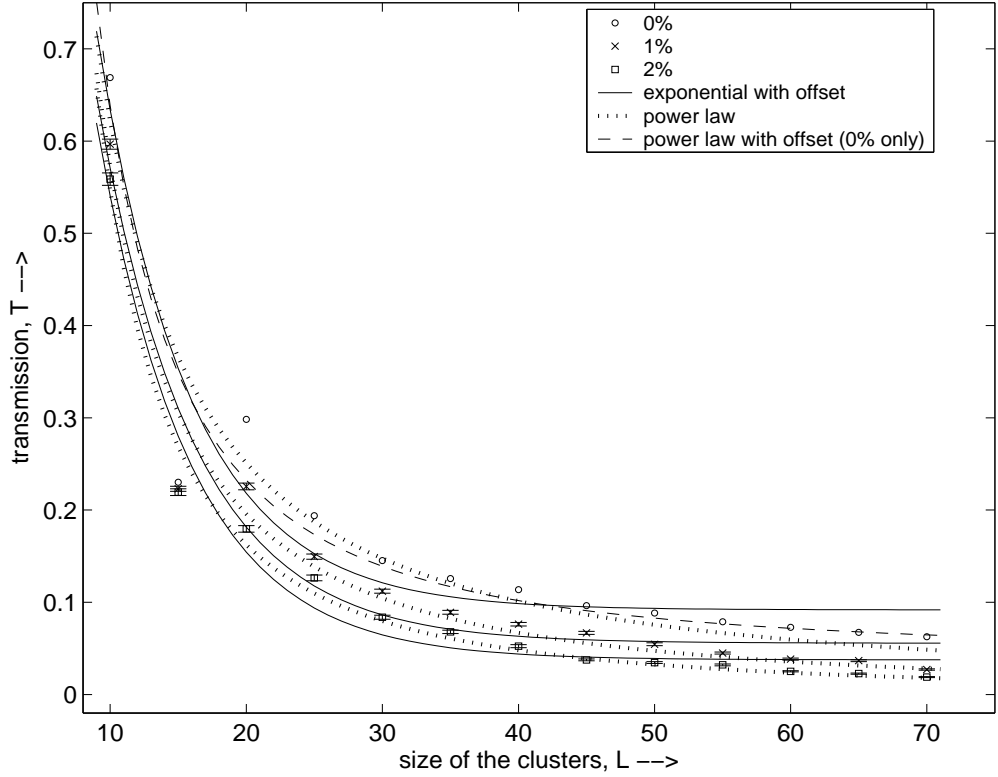}}}
\caption{The linear plot of the finite size scaling data of transmission through
         disordered clusters with off-diagonally connected leads for the lowest disorder values,
         all performed on the linear scale as direct, non-linear fits. For all three
         dilutions, 0\%, 1\%, and 2\%, the dotted and solid lines represent power-law
         fits and exponential fits with offsets, respectively, and in the order of 0\%,
         1\%, and 2\% from the top line to the bottom one for each type of fit. From
         this plot, the power-law fits appear to be better than the exponential with
         an offset in all cases unlike for the diagonal connection, though a power-law
         with offset may arguably fit better for the zero-disorder limit.}
\label{scal2b}
\end{figure}

More closely observing the results from the lowest values of disorder,
Fig.~\ref{scal2b} shows the data for the off-diagonal connection on the linear scale,
together with the results of power-law fit and exponential fit with a finite offset
performed as direct, non-linear fits on this scale for 2\% or less of disorder. In this
case a power-law fits the data significantly better than exponentials with finite offsets,
contrary to what we have observed in the case of diagonal connection. For 0\% disorder,
however, a power-law with finite offset appears to also fit the data well. Thus, it is evident
from Fig.~\ref{scal2b} and Table~\ref{tab4} that in this particular situation, our data
do not support the existence of finite $T$ for non-zero disorder in the thermodynamic
limit unlike the case of the diagonal connection.

We emphasize that this study measures transmission in the pseudo-one-dimensional system
with a 2D cluster sandwiched between 1D leads. Thus for the global transmission to occur
there has to be transmission through the 2D part of the system, requiring {\em extendedness}
for that part, but boundary perturbations where the chains connect to the cluster may
quench the transmission. While boundary effect cannot create global transmission if
there is none through the cluster, it could destroy it even if there is transmission through
the cluster portion. Though boundary effects must diminish as the cluster size increases,
this possibility seems inescapable because of the pseudo-one-dimensional nature of our system.
Therefore, the evident lack of global transmission for the case of off-diagonal connection
may be due to its quenching of a tenuous extendedness of the cluster due to a boundary perturbation.

\begin{table}[h]
\caption{Fitting parameters of non-linear fits for off-diagonal connection at low
disorder. Shown in the parentheses are the lower and upper bounds for 95\% confidence
level.}
\label{tab4}
\begin{center}
\begin{tabular}{||c|c|c|c|c|c|c|c|c|c||}   \hline
Disorder   & \multicolumn{4}{c|}{Power law fit: $T = a\cdot L^{-b}$} & \multicolumn{5}{c||}{Exponential fit with offset: $T = a\cdot\mbox{e}^{-bL} + c$} \\ \cline{2-10}
           & a               & b               & $|R|^{2}$& SSE & a               & b                & c                   &$|R|^{2}$& SSE   \\ \hline
   0\%     &12.7 (3.6, 21.7) &1.31 (1.04, 1.58)& 0.92    &0.024 &2.33 (0.5, 4.16) &0.15 (0.08, 0.22) &0.09 (0.05, 0.14)    &0.90     & 0.028 \\ \hline
   1\%     &20.2 (8.7, 31.7) &1.55 (1.33, 1.77)& 0.97    &0.009 &2.11 (0.91, 3.32)&0.14 (0.09, 0.19) &0.06 (0.02, 0.09)    &0.94     & 0.014 \\ \hline
   2\%     &30.9 (17.7, 44.1)&1.75 (1.58, 1.92)& 0.99    &0.003 &2.17 (1.23, 3.11)&0.15 (0.11, 0.19) &0.04  (0.02, 0.06)   &0.97     & 0.007 \\ \hline
\end{tabular}
\end{center}
\end{table}

Summarizing, our work clearly finds exponential localization at least for a
dilution of 20\% and greater for the energies and connection types treated here.
In addition, it also strongly supports the existence of a power-law localized regime
at lower disorder as suggested by many in the past. Thus, even though we have not
specifically searched for the point of transition between the two localized regimes, there must
obviously be a transition between them as we find both at different amounts of disorder.
Furthermore, our data are consistent with the existence of an extended state with
a finite $T$ in the thermodynamic limit, and thus also a delocalization transition
to it, at least for certain energies of the incident particle and for
certain types of lead connections (e.g., our diagonal, point-to-point connection),
at very low disorder amounts (1\% and 2\% disorder among our data). Though these findings
may not be entirely new individually, we believe that this is the first study in which
all three regimes are identified in the same system as the disorder amount is varied.

The issue of the localization length $L_l$ must however be discussed since
even if the system happens to be exponentially localized in the thermodynamic limit,
in order to observe it as such, the localization length of
the wave packet must be substantively smaller than the system size. Otherwise,
one may observe what appears to be a weaker form of localization (such as power-law)
or even an apparently extended state. So it would obviously be useful if we have a
theoretical estimate of $L_l$ to compare with the system size.

According to the standard approach for an electron transport,
$L_l \sim \mbox{exp}(k_f l_m)$ where $k_f$ is the Fermi wave number
of the electron and $l_m$ is its mean free path \cite{abrahams:79}. Assuming that
the mean free path is determined by scattering from the vacant sites, a crude
estimate of $\l_m$ is the mean separation between vacant sites. Denoting the
dilution by $q$ ($0 < q < 1$), this amounts to $l_m \sim 1/q^{1/2}$, or about 10
lattice spacings for 1\% dilution and 2 lattice spacings for 20\% dilution. The
estimation of what to use in place for $k_f$ is, however, much more difficult.
If we substitute the central wave number for the incident wave packet while it is
on the 1D input lead, say, $k_{0}$ = 4.7, into the exponential expression for $L_l$
together with the above crude estimates of $l_m$, the resulting values come out
to be much larger than any conceivable system size one can simulate
for almost any amount of dilution. So, according to this scenario, our data should
not show exponential localization at all. However, we clearly observe an
exponentially localized regime at least for dilutions of about 20\% or greater.
Thus, it must be that the relevant wave number for the particle as it spreads within
the 2D cluster region is rather smaller than $k_{0}$ on the 1D input chain.
Since we do not have an independent theoretical estimate of this wave number,
the deductive approach outlined above eludes us at this time.

Instead, we have taken a different approach here. Our finite size scaling approach
does not rely on any single system size but rather focuses on the trend as it
increases. In fact this analysis detects different trends depending on the dilution
amount and we can estimate the localization length from the data themselves
at least in the clearly observed exponentially localized regime. Indeed,
where Tables~\ref{tab1} and \ref{tab2} indicate the best fits to be the
exponential ($a\cdot\mbox{e}^{-bL}$), we can estimate $L_l \sim b^{-1}$, interpreting
the different system sizes $L$ as analogous to different length scales of
observation on a very large system. Table~\ref{tab3} summarizes the simulation
estimates of $L_l$ made in this way in the regions of disorder where the respective
exponential fits are significantly better than other types of fits. From these estimates,
it is evident that most of our system sizes at 15\% or more dilution are sufficiently large
compared with these estimates of $L_l$ and thus our results are internally consistent
with exponential localization at these amounts of disorder.

\begin{table}[h]
\caption{Estimated values of the localization lengths from $b^{-1}$ in fitting transmission to
an exponential without an offset}
\label{tab3}
\begin{center}
\begin{tabular}{||c|c|c||}   \hline
Disorder   & \multicolumn{2}{c||}{Localization length} \\ \hline
           & Diagonal    & Off-diagonal       \\
           & connection  & connection         \\  \hline
   15\%      &   -         & 12.2             \\  \hline
   20\%      &  8.3        & 8.6              \\  \hline
   25\%      &  6.4        & 6.5              \\  \hline
   30\%      &  4.7        & -                \\  \hline
\end{tabular}
\end{center}
\end{table}

We could, in addition, also force fits to an exponential without an offset for
smaller dilution where other fits such as power-laws or those with a non-zero
offset are actually significantly better than the pure exponential. This is
because, if in fact the state were exponentially localized but it happens to
fit better either a power-law or an exponential with a finite offset because
of the small system size, then one might expect a forced, pure-exponential fit
to reveal a hint of the localization length that is larger than the system size.
When this is done for the diagonal connection, the resulting estimates of $L_l$
vary as 10 (for 15\% disorder), 10 (10\%), 16 (5\%), 32 (2\%), 50 (1\%), and 111 (0\%),
while the off-diagonal connection data of Fig.~\ref{scal2a} vary as 17 (10\%), 18
(5\%), 20 (2\%), 22 (1\%), and 30 (0\%). Thus, the estimates for the diagonal
connection do not agree with such expectation, or, in other words, the system size
limitation does not seem to be a factor in the apparent existence of the extended
regime, at least for 1\% and 2\% disorder. Those for the off-diagonal connection do
not agree with the expectation as well for all values of dilution used, thus in this
case, the power-law localization appears to be genuine and not the artifact of the
system size limitation.

\section{Summary and Conclusion}

In this work we have developed a tool to visualize the propagation of a Gaussian
wave packet through a 2D cluster connected with 1D leads under various conditions.
We applied this dynamical method to study the behavior of a quantum particle in
completely ordered as well as disordered quantum percolation systems realized both
on the square and triangular lattices.
The method is very useful to study the scattering of a quantum particle by potential
barriers. Since we can visualize and track the wave packet in real time, we can
investigate how a particle propagates through a disordered system and how it is
trapped inside the cluster.

For completely ordered systems, transmission is dominated by the interference effect.
It also depends on the way the leads are connected to the cluster in complex ways.
Transmission is generally high for the diagonally connected leads except for those
wave vectors for which destructive interference gives rise to close-to-zero
transmission (reflection resonances). For the busbar type connection, generally low
transmission is observed because of large interference effects of the wave packet
with the edge on the input side of the cluster, but with significant transmission
peaks (resonances) at certain energies.

An important application of this method may be to probe the existence or absence
of a localized-delocalized transition in the two-dimensional, disordered system. Our
finite-size scaling results imply that there exists a phase transition between a
power-law localization and an exponential localization regimes. In addition to this,
our study further suggests that there may exist delocalized states at very low
disorder, consistent with the proposition that delocalization transition exists in the 2D
disordered quantum percolation system, adding to the increasing literature on
this still controversial issue. Of course, since the present work is based on
specifically constructed wave packets and is based on relatively small and limited
types of geometry, further studies are called for to unequivocally determine the
answers to these issues. In particular, direct solutions of the Schr\"{o}dinger equation
along the lines of Daboul et al\cite{daboul:00} are desirable, and are being
investigated.\cite{islam:07}

\section*{Acknowledgments}

We would like to thank A. Overhauser, Y. Lyanda-Geller, S. Savikhin, Goldenfeld and many others
for discussions and the Physics Computer Network (PCN) at Purdue University where
most of the numerical work was performed.\\

\end{document}